\documentclass[%
reprint,preprintnumbers,
 amsmath,amssymb,
 aps,superscriptaddress
]{revtex4-1}

\usepackage[hidelinks]{hyperref} 
\usepackage[dvipsnames]{xcolor}
\usepackage{colortbl}
\usepackage{subcaption} % for supplemental information only
\hypersetup{
    colorlinks,
    linkcolor={red!50!black},
    citecolor={blue!50!black},
    urlcolor={blue!80!black}
}
\usepackage{graphicx}% Include figure files
\usepackage{dcolumn}% Align table columns on decimal point
\usepackage{bm}% bold math
\usepackage{verbatim}
\usepackage{soul} % for strikethrough \st

\usepackage{changes}

\begin{document}

\title{Photocatalytic water oxidation on SrTiO$_3$ [001] surfaces}
\author{Vidushi Sharma}
\affiliation{Theoretical Division, Los Alamos National Laboratory, Los Alamos, NM 87545, USA}
\affiliation{Center for Nonlinear Studies (CNLS), Los Alamos National Laboratory, Los Alamos, NM 87545, USA}
\affiliation{Department of Physics and Astronomy, Stony Brook University, Stony Brook, NY 11794-3800, USA}
\affiliation{Institute for Advanced Computational Science, Stony Brook University, Stony Brook, NY 11794-3800, USA}
\author{Benjamin Bein}
\affiliation{Department of Physics and Astronomy, Stony Brook University, Stony Brook, NY 11794-3800, USA}
\author{Amanda Lai}
\affiliation{Department of Physics and Astronomy, Stony Brook University, Stony Brook, NY 11794-3800, USA}
\author{Bet\"{u}l Pamuk}
\affiliation{School of Applied and Engineering Physics, Cornell University, Ithaca, NY 14853, USA}
\author{Cyrus E. Dreyer}
\affiliation{Department of Physics and Astronomy, Stony Brook University, Stony Brook, NY 11794-3800, USA}
\affiliation{Center for Computational Quantum Physics, Flatiron Institute, 162 5$^{th}$ Avenue, New York, NY 10010, USA}
\author{Marivi Fern\'{a}ndez-Serra}
\email{maria.fernandez-serra@stonybrook.edu}
\affiliation{Department of Physics and Astronomy, Stony Brook University, Stony Brook, NY 11794-3800, USA}
\affiliation{Institute for Advanced Computational Science, Stony Brook University, Stony Brook, NY 11794-3800, USA}

\author{Matthew Dawber}
\email{matthew.dawber@stonybrook.edu}
\affiliation{Department of Physics and Astronomy, Stony Brook University, Stony Brook, NY 11794-3800, USA}

\begin{abstract}
%\begin{linenumbers}
SrTiO$_3$ is a highly efficient photocatalyst for the overall water splitting reaction under UV irradiation.
However, an atomic-level understanding of the active surface sites responsible for the oxidation and reduction reactions is still lacking. 
Here we present a unified experimental and computational account of the photocatalytic activity at the SrO- and TiO$_2$- terminations of aqueous-solvated [001] SrTiO$_3$.
Our experimental findings show that the overall water-splitting reaction proceeds on the SrTiO$_3$ surface only when the two terminations are simultaneously exposed to water.
Our simulations explain this, showing that the photogenerated hole-driven oxidation primarily occurs at SrO surfaces in a sequence of four single hole transfer reactions, while the TiO$_2$ termination effects the crucial band alignment of the photocatalyst relative to the water oxidation potential.
%Our simulations explain this, showing that the photogenerated-hole driven oxidation primarily occurs at SrO surfaces in a sequence of four single hole transfer reactions. 
%
%We demonstrate that while such reactions are preferentially catalyzed at the SrO surface, the TiO$_2$ termination effects the crucial band alignment of the photocatalyst relative to the water oxidation potential.
%
The present work elucidates the interdependence of the two chemical terminations of SrTiO$_3$ surfaces, and has consequent implications for maximizing sustainable solar-driven water splitting.

%\end{linenumbers}
\end{abstract}

%Use showkeys class option if keyword
                              
\preprint{LA-UR-22-20654}
                              
\maketitle

%\tableofcontents
\section{Introduction}
Photocatalytic water splitting is a promising route to decrease our energy dependence on fossil fuels \cite{Kudo2009,Maeda2007}.
Redox-active oxides like TiO$_2$ are ideal material platforms to
study and optimize the heterogeneous oxidation and reduction reactions to convert water into H$_2$
and O$_2$ using solar photons as the sole source of energy \cite{Selloni2020, Linsebigler1995, Eidsvag2021, Miyoshi2018}.
Some of the best photocatalysts are oxide materials \cite{Rajeshwar2007, Lee2021} and specifically perovskite oxides \cite{Kanhere2014}.
Perovskite oxide materials have the ability to selectively separate the photogenerated electrons and holes and efficiently transfer them to the semiconductor surfaces, where they
can drive the two redox half reactions \cite{Li2013}.
SrTiO$_3$ is a prototypical cubic perovskite and was first proposed as photocatalyst for water splitting to generate hydrogen in 1976 \cite{Wrighton1976}. 
Despite having a band gap of 3.25 eV \cite{vanBenthem2001} which restricts the photons absorbed to the ultraviolet range of the solar spectrum, SrTiO$_3$ is a well-studied photo-reactive material and serves as a platform for understanding photocatalytic water splitting in more complex systems \cite{Wrighton1976, Giocondi2003, Kato2013, Kato2002, Wang2006, Wei2009, Niishiro2005, Konta2004, Ohsawa2006, Tanaka2010, Wang2009, Subramanian2006}.
Thus many different factors  influencing the quantum efficiency of SrTiO$_3$ have been investigated.
These include, for example, the effects of doping \cite{Asai2014, Furuhashi2013, Jiang2020}, the influence of different facets \cite{Kato2013, Zhu2016}, and the effects of the pH of the solution \cite{Zhang2020}. 
%
%\CD{I like this lead up, but then there is a detour into some details about the reaction. Added a few sentences here to try and bring the point home early, borrowing from other parts of the introduction. Please check.} 
%
However, a complete microscopic picture of the photocatalytic process, even in this model system, is missing.
A significant reason for this is that many aspects of the process depend on the details of the SrTiO$_3$ surface, and its aqueous interface. Considerable work has been performed in recent years on understanding the photocatalytic nature of these surfaces \cite{Zhu2016, Zhang2020, Giocondi2003, Schultz2011, Ham2016}, but far less is known about the atomistic details of the oxidation mechanism at the aqueous interface \cite{Shen2010}.
In part this is due to a lack of experimental data obtained from samples designed with controlled surface properties.
Such  information would be critical in integrating theoretical models and experiments in a unified study \cite{Lee2021}.

The oxidation reaction of water to molecular oxygen is a complicated four-electron reaction, coupled to the reduction of water into molecular hydrogen.
In heterogeneous photocatalysis, the photogenerated holes drive the water oxidation at the semiconductor surface:
\begin{equation}
\label{eq:ox}
 2\textrm{H}_{2}\textrm{O}_{(\textrm{aq})} + 4 \textrm{h}^+ \rightarrow \textrm{O}_{2(\textrm{g})} + 4 \textrm{H}^+_{(\textrm{aq})} ~.
\end{equation}
In an overall water-splitting material, the hydrogen production or proton reduction also occurs  at the surface:
\begin{equation}
\label{eq:red}
\textrm{ 2H}^+_{(\textrm{aq})} + 2\textrm{e}^- \rightarrow \textrm{H}_{2(\textrm{g})} ~.
\end{equation}
The two half reactions need to proceed at the same rate or else the full redox reaction will be shut down by charge accumulation.
Hence among the factors limiting the quantum efficiency, charge separation is a dominant one \cite{Li2013, Zhu2016, Zhang2020, Selcuk2016}.
This means that the photogenerated electron hole pairs, which in SrTiO$_3$ usually exist in the form of a self-trapped exciton or  separated electron and hole polarons \cite{Crespillo2019}
need to arrive at their corresponding surface reaction active sites at similar rates in order to maximize the efficiency of the overall water splitting reaction.
This idea was already proposed by Zhang {\it et al.} \cite{Zhang2020},  in their study of photochemical reactivity at different SrTiO$_3$ surfaces.

In this work, we identify the explicit role that surface chemistry and termination have in the overall water splitting reaction in SrTiO$_3$ [001] surfaces, using a combination of experimental and computational approaches.
%
%The goal of this work is to provide a combined experimental/theoretical understanding of the water oxidation reaction in sample perovskite oxide surfaces, where the surface termination  experimentally controlled.
%Using  a combination of treatment and growth techniques we
%can control the termination of the water exposed surfaces, allowing us to reliably couple the theoretical and experimental aspects, providing an unprecedented insight into the photochemical activity at aqueous
%SrTiO$_3$ interface.
%
%This degree of control allows us to reliably couple the theoretical and experimental aspects, providing an unprecedented insight into the photochemical activity at aqueous
%SrTiO$_3$ interface.
%
Crucially, our surface treatment techniques allow us to deterministically produce samples with only SrO termination, only TiO$_2$ termination, or a mixture of both.

%Experimentally we use surface preparation and characterization techniques to identify how the overall oxidation reaction proceeds in SrTiO$_3$ [100] surfaces as a function of the two -- SrO or TiO$_2$ -- possible terminations. \CD{I would think this is a unique aspect (at least at the level of studying photocatalysis)? If so, we should emphasize it. Something like: ``Crucially, our surface treatment techniques allow us to deterministically produce samples with only SrO termination, only TiO$_2$ termination, or a mixture of both.}
%
To experimentally evaluate the redox reaction, we replace Eq.~\eqref{eq:red} by the reduction of an electron scavenger. Specifically, we observe the reduction of Ag$^+$ into metallic Ag; thus the detection of metallic Ag at the surface of the semiconductor serves as a proxy for the evaluation of the efficiency of the overall redox reaction \cite{Giocondi2003, Schultz2011, Zhang2020}. %\CD{Sounds like from the references, this has been done before? Otherwise we can emphasize it also.}
These experiments demonstrate that SrTiO$_3$ is only photocatalytically active if \emph{both} SrO and TiO$_2$ terminations are present on the surface.
A microscopic explanation for this result is given by \textit{ab initio} molecular dynamics simulations (AIMD) of water atop an SrTiO$_3$ slab with the two relevant surfaces exposed to water.
In addition to providing a clear atomistic description of these interfaces, the simulation results allow us to evaluate and propose a model for the oxidation reaction at the surface that explains the experimental results.
Even more, the combined results provide information
on the nature of the photo-excited carriers, by identifying the spatial correlation between the oxidation and reduction sites at the surfaces.
%\CD{I added a bit more here about what was found, please check.}

%In the first part of the paper we investigate the photocatalytic water oxidation on SrTiO$_3$ \st{thin film samples} [001] surfaces, focusing on how the interplay of the two -- SrO or TiO$_2$ -- surface terminations influences the overall reaction.
%
%We also perform \textit{ab initio} molecular dynamics simulations (AIMD) of water atop a slab of SrTiO$_3$ with the two relevant surfaces exposed to water.
%

\section{Photocatalysis on engineered $\mathbf{SrTiO_3}$ [001] surfaces}

The focus of this work is characterizing photocatalysis on the [001] surfaces of SrTiO$_3$. 
Along the (001) direction the material can be viewed as a stack of alternating TiO$_2$ and SrO planes, and  [001] surfaces can be terminated  in either of these two planes. 
Practically, all substrates cut from a crystal will have a small miscut angle so that the surface will not be an exact [001] plane. 
In the absence of further treatment, the surface of a substrate that has been cut and polished will present a mixture of the two possible terminations.
However, there are procedures to obtain atomically-flat surfaces with a well-defined step and terrace structure, typically used as preparation for the growth of epitaxial thin films \cite{kawasaki1994atomic,koster1998quasi,ohnishi2004preparation}. 
We  leverage these techniques to provide a controlled surface for the study of photocatalysis.
%
%These can be used to prepare well defined surfaces to help understand the surface photocatalytic processes from an atomistic point of view.
%
%To achieve this control, in this work, we have engineered surfaces which either present a single termination, or separated SrO and TiO$_2$ into well defined arrangements.
%
%Atomic force microscopy allows us to evaluate the quality and structural characteristics of these prepared surfaces.
%
%We then use the previously described Ag photo-deposition proxy reaction to determine the active sites of reaction on the treated surfaces.
%
 
%
The substrate vendor (Cyrstec) provided us with two type of substrates.
(i) untreated  substrates, which are just cut and polished  and (ii) single termination substrates  substrates that had been treated to achieve a TiO$_{2}$ termination \cite{koster1998quasi}. 
Untreated substrates were subsequently etched, via high-temperature high-pressure water etching \cite{Velasco2013}. 
The water etching method has an advantage over treatments that involve buffered HF in that it avoids F$^-$ impurities which can be introduced by traditional buffered hydrofluoric acid etching \cite{Chambers2012}.
The water treatment leads to mostly pure TiO$_2$ surfaces with patches of SrO.
Subsequent annealing can then be used to modify the SrO coverage of the substrates \cite{Bachelet2009}.

\begin{figure}[h!]
    \includegraphics[width=8.5cm]{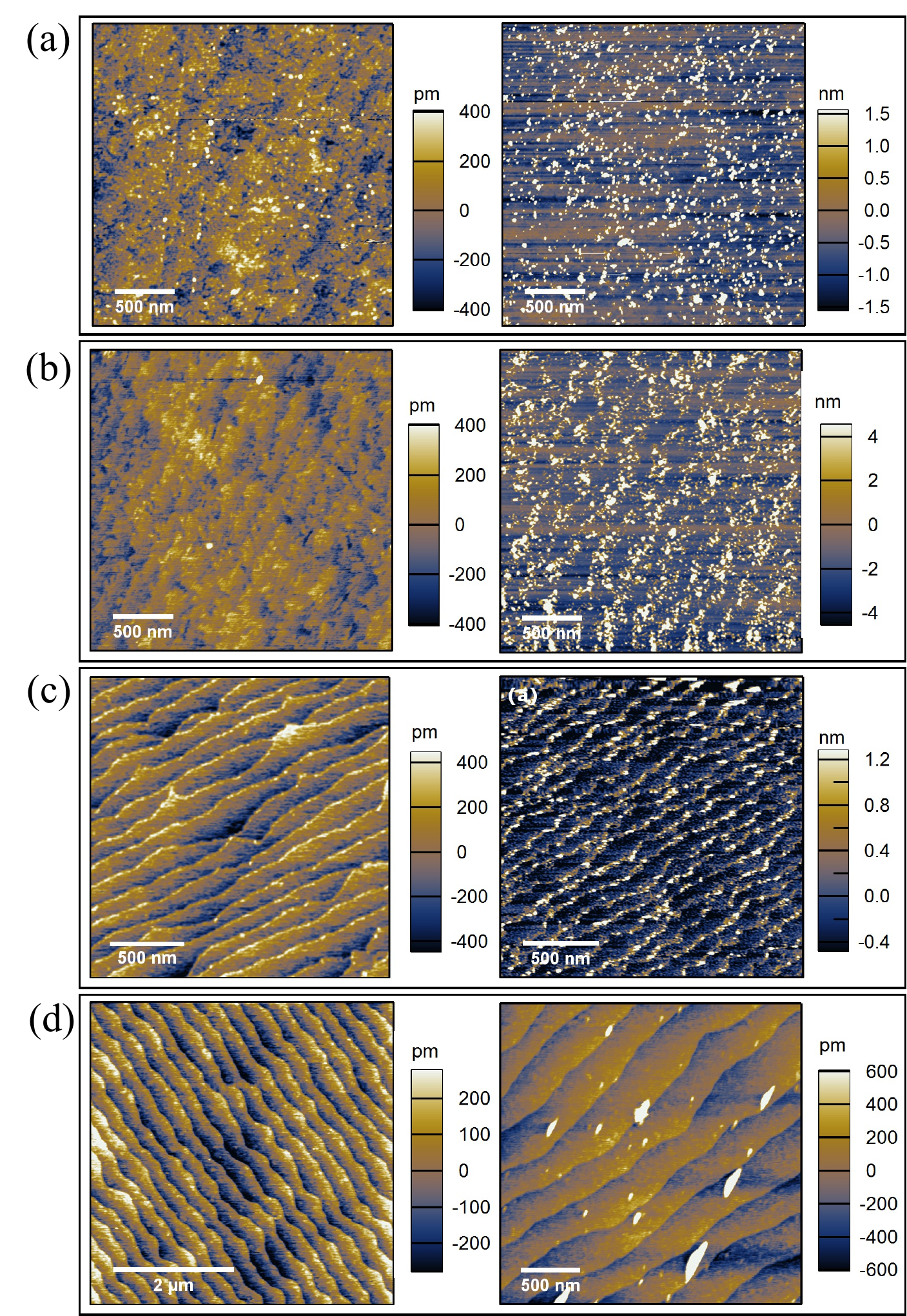}
    \caption{AFM topography scans of surfaces before, left, and after, right, silver deposition. (a) SrTiO$_{3}$ surface that has been cut and polished only, (b) surface that has been treated with a procedure that only partially segregates the SrO on the surface, (c) surface that has been treated such that SrO is segregated to the step edges, (d) surface that has been treated to obtain SrO termination by extending the duration of the high temperature anneal.}
    \label{fig:fig1}
\end{figure}

Photocatalytic activity on all surfaces is evaluated using the Ag$^+$ proxy reduction method for water splitting developed in \cite{Giocondi2003, Kudo2009, Tiwari2009, Kato2013}.
Upon UV illumination, water is oxidized to O$_2$ and H$^+$ (see Eq.~\eqref{eq:ox}) while the reduction of H$^+$ (Eq.~\eqref{eq:red}) is replaced with the reduction of Ag$^+$ ions, resulting in Ag deposited on the surface.
%
% \begin{equation}
% \label{eq:Ag}
% \textrm{(Proxy-reaction):  Ag}^+_{(\textrm{aq})} + \textrm{e}^- \rightarrow \textrm{Ag}_{(\textrm{s})}
% \end{equation}
%
The photocatalytically deposited silver can then be observed using atomic force microscopy (AFM). As explained in the Methods section, our standard experiment is to expose the sample to UV illumination for 5  minutes while it is placed in a AgNO$_{3}$ solution.
When this experiment is performed on an untreated surface (Fig.~\ref{fig:fig1} (a, left)) the silver appears to be deposited fairly uniformly (Fig.~\ref{fig:fig1} (a, right)), with no particular order visible.
Fig.~\ref{fig:fig1} (b, left) shows a sample that was treated using a high pressure water etch, annealed at  650$^{\circ}$C for 24 hrs, retreated, and then annealed again for 24h at 750$^{\circ}$C.
The step edges are now visible in the etched sample, but are not very straight.
The thermal treatment of this sample has led to residual SrO starting to diffuse towards the step edges.
%\CD{Can you say how we know this?}
%
Photocatalyzed silver deposition appears to be associated with the residual SrO, as observed in Fig.~\ref{fig:fig1} (b, right).

\begin{figure}[h!]

    \includegraphics[width=8.5cm]{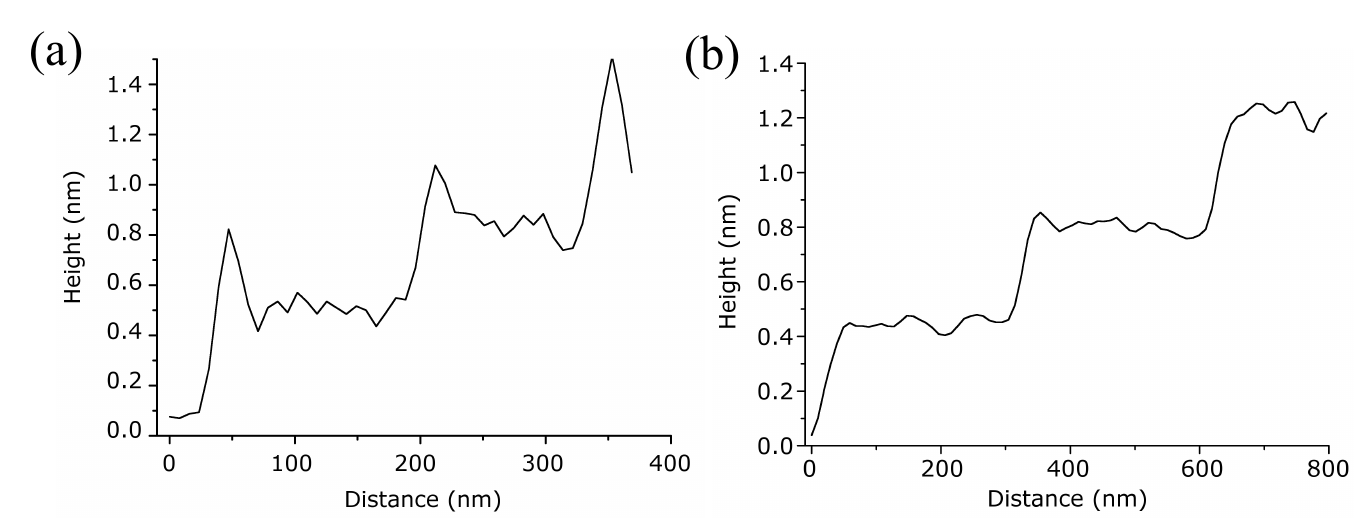}
    \includegraphics[width=8.5cm]{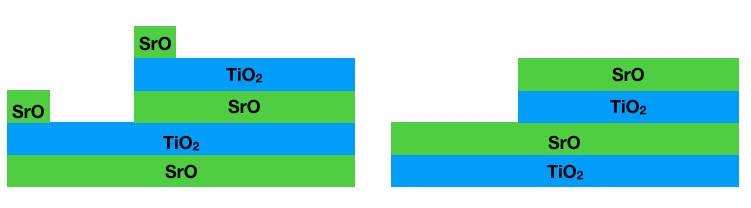}
    \caption{AFM line profiles perpendicular to step edges.  (a) Line profile of a surface that has been treated such that SrO is segregated to the step edges. (b) Line profile of a surface that has been treated to obtain SrO termination.
    Bottom figures illustrate the distribution of SrO and TiO$_2$ planes corresponding to the line profiles. }
    \label{fig:fig2}
\end{figure}

That indeed Ag deposition seems to occur in the vicinity of SrO patches
is more evident in our next treated sample shown in Fig.~\ref{fig:fig1} (c).
This sample is obtained after adding an additional 2hr anneal at 900$^{\circ}$C.
This results in the SrO completing its diffusion along the TiO$_{2}$ step and collecting on the step edge as a 1/2 unit cell layer (Fig. \ref{fig:fig1} (c, left)). 
This is most obvious in a line profile taken perpendicular to the step edges (Fig. \ref{fig:fig2} (a)).
Here it can be seen that there is a 0.6 nm jump at each step edge which corresponds to 1.5 unit cell steps, the change from SrO to TiO$_2$ termination happens with a 0.2 nm drop which corresponds to 0.5 unit cell steps. 
The preference of SrO to segregate along the step edges is a known property \cite{szot1999surfaces} and a similar surface was previously achieved by Bachelet \textit{et al.}, albeit with a different heat treatment \cite{Bachelet2009,bachelet2009self}.
After photo-reactivity on this surface, the silver is clearly found in the vicinity of the step edges where the SrO has collected (see Fig.\ref{fig:fig1}(c) and Fig.\ref{fig:fig2}(a)).

A fully SrO terminated surface can be obtained by replacing the anneal after the second treatment with a sustained high temperature anneal (38hrs at 900$^{\circ}$C). 
This extended high temperature anneal causes SrO to vertically diffuse from the bulk of the sample \cite{szot1999surfaces} and produces a highly ordered single termination SrO surface (Fig.~\ref{fig:fig1} (d, left)).
As shown in Fig.~\ref{fig:fig2} (b)  this surface has single unit cell transitions at the step edges.
Remarkably, as observed in Fig.~\ref{fig:fig1} (d, right) little silver is deposited on this surface, indicating that a pure SrO surface is not photocatalytically active.

\begin{figure}[h!]
    \includegraphics[width=8.5cm]{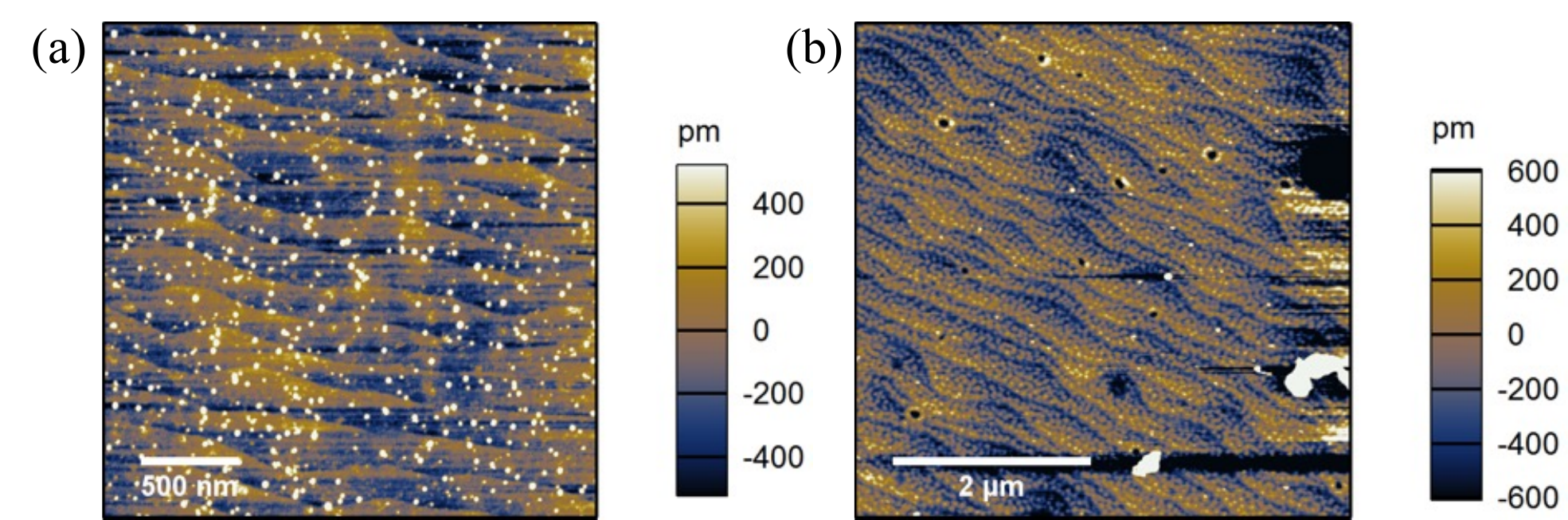}
    \caption{AFM topography scans of TiO$_{2}$ terminated surfaces after silver deposition(a) treated substrate (b) deposited film.}
    \label{fig:figtio2}
\end{figure}

We also carried out the silver deposition procedure on two TiO$_{2}$ terminated surfaces.
The first is on a TiO$_2$-substrate  treated by the vendor with a buffered HF etch and high temperature anneal. 
Here we observe some silver being deposited, (Fig. \ref{fig:figtio2} (a)) but there is no apparent correlation between the deposition sites and the surface morphology of the film. We associate this small amount of Ag deposited to the presence of F$^-$ impurities at the surface.

For the second surface,  we grew SrTiO$_3$ thin films using an off-axis RF magnetron sputtering  on top of vendor-SrTiO$_3$ substrates with TiO$_{2}$ termination. It is expected that these films will maintain a TiO$_{2}$ termination.
After performing a photocatalytic silver deposition very few silver particles were formed on these surfaces, even if we extend the exposure time to 1 hour, as shown in Fig. \ref{fig:figtio2} (b).

Hence our experiments show that (i) single terminated surfaces (SrO or TiO$_2$) are not photocatatytically active, and
(ii) mixed terminated surfaces are active, and Ag is deposited near SrO terminations. 

\section{Atomistic structure of solvated S\lowercase{r}T\lowercase{i}O$_3$ (001) surfaces}

In order to obtain an atomistic picture of the photocatalytic process, we performed \textit{ab initio} molecular dynamics simulations (AIMD) of a SrTiO$_3$
slab in the presence of liquid water. The slab was 3 unit cells along the (001) direction and $2\sqrt{2}\times2\sqrt{2}$ unit cells
along the in-plane (110) direction.
Since we choose to have the slab contain an integer number of unit cells, it is necessarily terminated by a SrO surface on one side, and a TiO$_2$ surface on the other.
This choice allows us to explore in one single simulation the structure of the two different terminations.
We add 64 water molecules between the periodically-repeated slabs, which we confirmed were enough to screen the two different potential offsets arising from this asymmetric slab. 

 \begin{figure}[h]
 \includegraphics[width=3.in]{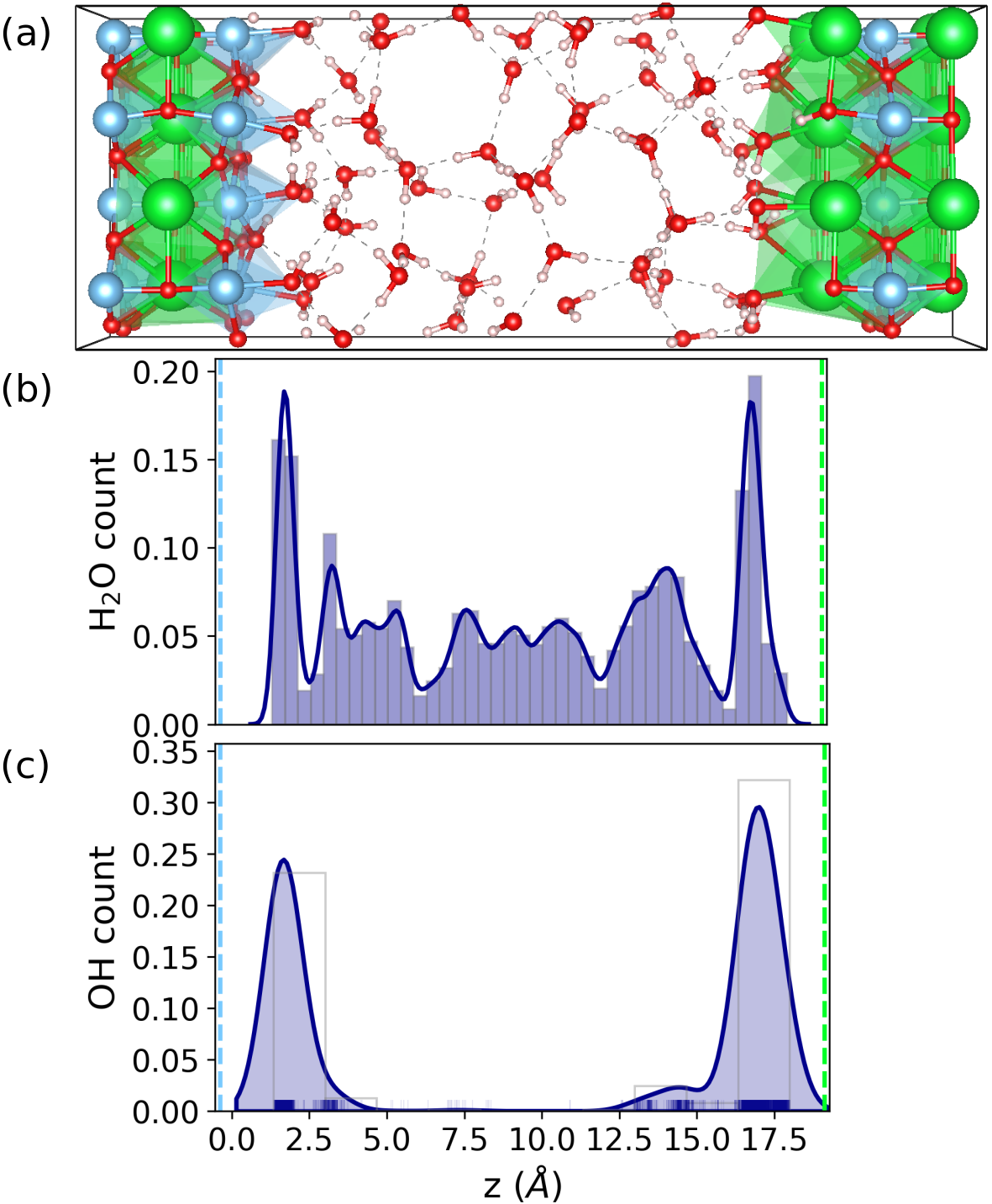}
 \caption{From an equilibrated molecular dynamics simulation: (a) $(001)$ SrTiO$_3$ surfaces -- TiO$_2$ (left, in blue) and SrO (right, in green) -- interacting with a box of 64 water molecules. The number density distribution along the vertical $z-$direction for: (b) H$_2$O molecules and, (c) dissociated OH$^-$ species. The positions of TiO$_2$-- and SrO-- surfaces with respect to the box of water are shown in blue and green dashed lines respectively.}
 \label{fig:sto_waterfull}
 \end{figure}
 
Fig.~\ref{fig:sto_waterfull} shows a snapshot of the simulated system.
The two surfaces quickly dissociate water and achieve an equilibrium state within 10 ps of simulation. 
Experimentally, the presence of OH$^-$ at the [001] SrTiO$_3$ surface
has been thoroughly studied by Domingo {\it et al.} \cite{domingo2019water}.
As seen in Fig.~\ref{fig:sto_waterfull}(c), the two surface terminations are active in dissociating water into OH$^-$ (which binds to Sr/Ti depending on the surface termination)  and H$^+$ (which always binds to surface O atoms).
However, the SrO terminated surface is more effective than TiO$_2$ in this task.
SrO presents 35$\%$ of adsorbed and dissociated water molecules versus 25$\%$ for TiO$_2$. 
In addition, we observe secondary dissociation events at the SrO surface, which cause the additional peak shoulder in Fig.~\ref{fig:sto_waterfull}(c) (right).
These are transient proton transfer reactions between surface OH$^-$ and nearby
H$_2$O molecules, indicating that this surface has a lower Pka or higher surface acidity \cite{Wang2012} than a TiO$_2$ termination. 
%
%This indicates that the pH near the two surfaces is rather different.
%

\section{Electronic structure and Band alignment}

\begin{figure}[h]
    \centering %remove centering command for the paper. 
    \includegraphics[width=2.5in]{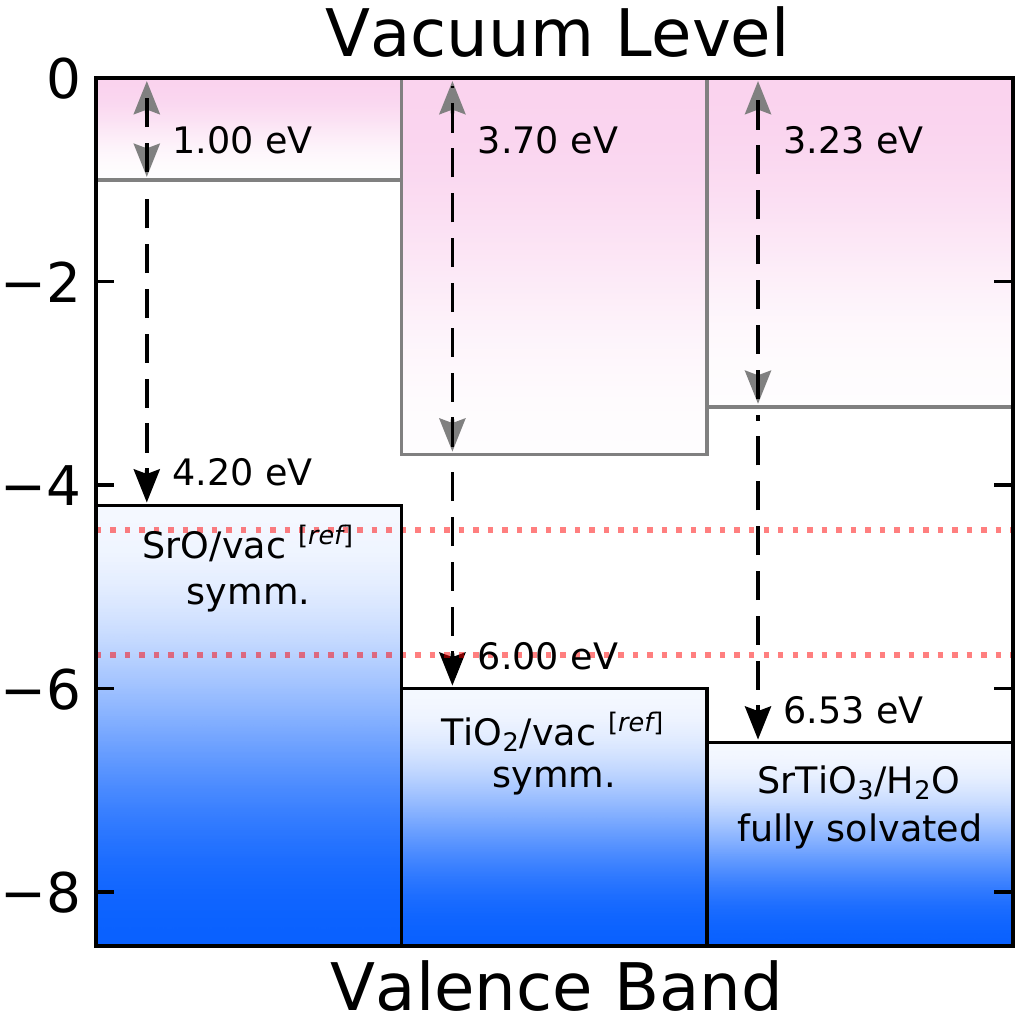}
    \caption{Band alignments for symmetric SrO-terminated, symmetric TiO$_2$-terminated SrTiO$_3$/vacuum [ref]=\cite{Ma2020}, and fully-solvated SrTiO$_3$/H$_2$O slabs using DFT HSE06. The red dotted lines indicate the water redox potentials referenced to the vacuum level (E$_{\text{H}^+/\text{H}_2} = -4.44$ eV, E$_{\text{O}_2/\text{H}_2\text{O}} = -5.67$ eV).}
    \label{fig:bandalignment}
\end{figure}

Among the many factors that control the photocatalytic efficiency of a semiconductor surface, the relative alignment of the semiconductor band edge and the corresponding redox level in water determines
whether the photo-excited carriers can carry the oxidation and reduction reactions.
In our case, the reduction half reaction is bypassed by the reduction of Ag$^+$ and we seek to understand the oxidation reaction, and in particular, its dependence on the semiconductor surface termination. 
To this end, we must obtain the alignment of the valence band edge (VBE) with respect to the electrochemical water oxidation potential \cite{Shen2010,Kharche2014} for each surface termination.
%
%We follow the procedure described by Kharche {\it et al.}\cite{Kharche2014} to treat the level alignment at aqueous semiconductor interfaces.
%
In oxide perovskites (ABO$_3$), the work function difference  between the AO- and BO$_2$-terminated (001) surfaces is theoretically predicted to be on the scale of a few eV \cite{Zhong2016, Sung2020}.
While the specific case of pure SrTiO$_3$ has been studied in detail using density functional theory (DFT)-based methods \cite{Ma2020, Marzari2020, Kim2015}, the aqueous interface band alignment
remains unexplored.
Here we evaluate the electronic structure on samples of the previously described STO slab, after dissociation at the two water exposed surfaces reaches equilibrium.
In the procedure proposed by Kharche {\it et al.} \cite{Kharche2014}, the alignment with respect to vacuum for the solvated slab is done using the 1$b_1$ level of the bulk water region of the simulated system as a reference.
This level is itself aligned with the 1$b_1$ level obtained from an independent water/vacuum slab calculation.
In Fig. \ref{fig:bandalignment} we present the vacuum-aligned band edge positions of a fully solvated asymmetric SrTiO$_3$ slab using a hybrid functional HSE06 \cite{HSE1, HSE2}. The first two columns show the band edges of pure SrO and TiO$_2$ terminated surfaces with respect to vacuum, as reported by Ma {\it et al.} \cite{Ma2020}, which we use as a reference for nonsolvated systems.
Additional calculations and discussion about the dependence of the results on the choice of exchange and correlation functional are presented in the methods and Supplementary Information. 
The water redox potentials are shown as red dotted lines.
As seen in this figure, a pure SrO surface is not favorable for water splitting, given that the corresponding VBE
is less positive (that is, closer to vacuum) than the water oxidation potential by $\sim 1.5$ eV.
On the other hand, pure TiO$_2$ surfaces present a VBE sufficiently positive to catalyze the water oxidation reaction, albeit with a small $\sim$ 0.3 eV overpotential \cite{Ma2020}. 
%
%We also present the level alignments as obtained for an ``asymmetric" system with an SrO termination on one side and a TiO$_2$ on the other, similar to our solvated system but with more layers along the non periodic direction.
%
%In this case, each surface has different alignment, with respect to the vacuum.
%
%The band edges on the two sides are calculated by using a layer-projected density of states.
% 
%The results are very similar to those obtained for the two symmetric systems. 
%
%However, we also present them here because they help to understand the results for the solvated interfaces.
%
%
%REMOVE U FROM FIG 6 AND PUT IT IN THE CAPTION
% \begin{figure*}
%     \centering %remove centering command for the paper. 
%     %\includegraphics[width=6.75in]{pcet_full.jpg}
%     \includegraphics[width=5.25in]{alignment.pdf}
%     \caption{Band alignments for SrTiO$_3$/vacuum, SrTiO$_3$/H$_2$O monolayer and fully-solvated at DFT+U; U = 4.45 eV with a QZP basis set for oxygen. The red dotted lines indicate the water redox potentials referenced to the vacuum level (E$_{\text{H}^+/\text{H}_2} = -4.44$ eV, E$_{\text{O}_2/\text{H}_2\text{O}} = -5.67$ eV).}
%     \label{fig:bandalignment}
% \end{figure*}
%ABOVE WAS REPLACED (MOVED TO SI) BY THE FOLLOWING FIGURE ON 8-NOV-2021

%{Pure SrO is the one that is catalytically active; surfaces present within how many nanometers -- (vs) width of the step size? important point experimentally length scale of the surfaces and how close they need to be for the redox reaction to take place.}
%TOO MUCH SRO PUSHES THE LEVELS UP SO WE NEED MORE TIO2 THAN SRO FOR GOOD YIELDS.
%
The fully solvated SrTiO$_3$ slab used to compute the alignment in Fig. \ref{fig:bandalignment} has one of each, i.e. SrO and TiO$_2$ surfaces exposed to water.
Hence we refer to this system as a 50$\%$ mixed surface slab.
%
%As explained in the methods section, the bulk water region mean $1b_1$ orbital energy has been chosen as our reference level and it is the same for both sides of the slab, that is, there is a single work function for both  surface terminations. 
%
%This would not be the case for an equivalent slab in vacuum, where each surface wouldhave a different alignment with respect to vacuum \cite{Sung2020}.
%
Although in vacuum each surface has a different work function, in aqueous solution, the surface water screens completely and within a very short distance (less than 5 \AA) the surface dipole due to the other termination, as further discussed in the supplementary information.
As a result of the screening that takes place through both dissociation and structural orientation, the band bending at the two surfaces results in more positive VBEs, placing them both $\sim 0.85$ eV below the water oxidation potential.

These results indicate that water dissociation, which induces a negative dipole moment (i.e. pointing into the surface) helps with the favorable level alignment for the overall water oxidation reaction at SrTiO$_3$ surfaces. 
However, the computed energy alignment upon solvation, which places the VBE of the solvated 50$\%$ termination slab 2.3 eV below that of the pure SrO unsolvated slab cannot be explained by the induced water screening dipole alone.
It was previously observed in \cite{Kharche2014} that this screening could account for an energy lowering of at most 0.5-1.0 eV.
In order to achieve an energy alignment sufficient to drive the oxidation reaction it is also necessary for both terminations to be present at the surface. 
This is because the surface dipole of SrO terminated surfaces is coupled to the corresponding dipole of TiO$_2$ terminations.
%
%Mixed surface terminations present a band offset that causes charge transfer to occur from SrO terminations to TiO$_2$ .
%
The overall VBE lowering will depend on the ratio
of one termination to another.
Our simulation results are obtained for a 50$\%$ ratio, but for different ratios, some bowing should be expected \cite{PhysRevB.83.134202}.
This is why pure SrO surfaces cannot drive the oxidation reaction and hence no Ag deposition is observed in the experiments.
However, these results are still not adequate to explain why pure TiO$_2$ surfaces are not photocatalytically active, nor do they explain why Ag is deposited at or near SrO terminations.
Therefore we augment our theoretical study with additional insights about the water oxidation reaction.

%Our AIMD provide us with sufficient atomistic information to explore the oxidation free energies for water in the two terminations.
%

\section{Photocatalytic water oxidation reactions}

Water oxidation on semiconducting surfaces can occur via a sequential four-step proton-coupled electron transfer (PCET) mechanism \cite{Shen2010,Hammes-Schiffer2021}.
Here we compute
the  free energy changes of four PCET reactions both at the SrO/water and TiO$_2$/water interfaces.
The proposed cycle intermediates match the homogeneous reactions for water oxidation in aqueous solutions.
%
%and TiO$_2$ surfaces the valence band edge is less positive (that is closer to vacuum)  than the water oxidation potential, hence in principle none of these surfaces are favorable for overall water splitting.
%\subsection{Water Oxidation reaction at SrO and TiO$_2$ terminations}
%In the realm of atomistic modeling of photocatalytically-active reactions, proton-coupled electron transfer (PCET) has emerged as a highly reliable mechanism for studying redox processes \cite{Hammes-Schiffer2021}.
%
%For example, it was shown by Shen {\it et al.} \cite{Shen2010} that water oxidation on GaN $(10\bar{1}0)$ surfaces can occur via a sequential four-step PCET mechanism.
%
%The standard oxidation free energies computed using a cluster model of this GaN/water interface suggested a favorable water oxidation process carried out by four photons in the ultraviolet (UV) range.
%
% proton-coupled electron transfer (PCET) mechanism.
%Using PCET to describe catalytic processes implies that each electron transfer occurs in concert with a proton transfer resulting in a lower overall activation barrier.
%
%The proposed cycle intermediates the homogeneous reactions for water oxidation in aqueous solutions.
%
%Here we follow the same approach, by studying the free energy changes of the same four PCET reactions both at the SrO/water and TiO$_2$/water interfaces.
%
At each of these steps, an incident UV photon generates an electron-hole pair. 
The resulting hole participates in the oxidation of water at an active surface site.
These reactions are coupled to the reduction of four Ag$^+$ ions at the surface by the corresponding electrons.

\begin{figure*}
    \centering %remove centering command for the paper. 
    \includegraphics[width=6.2in]{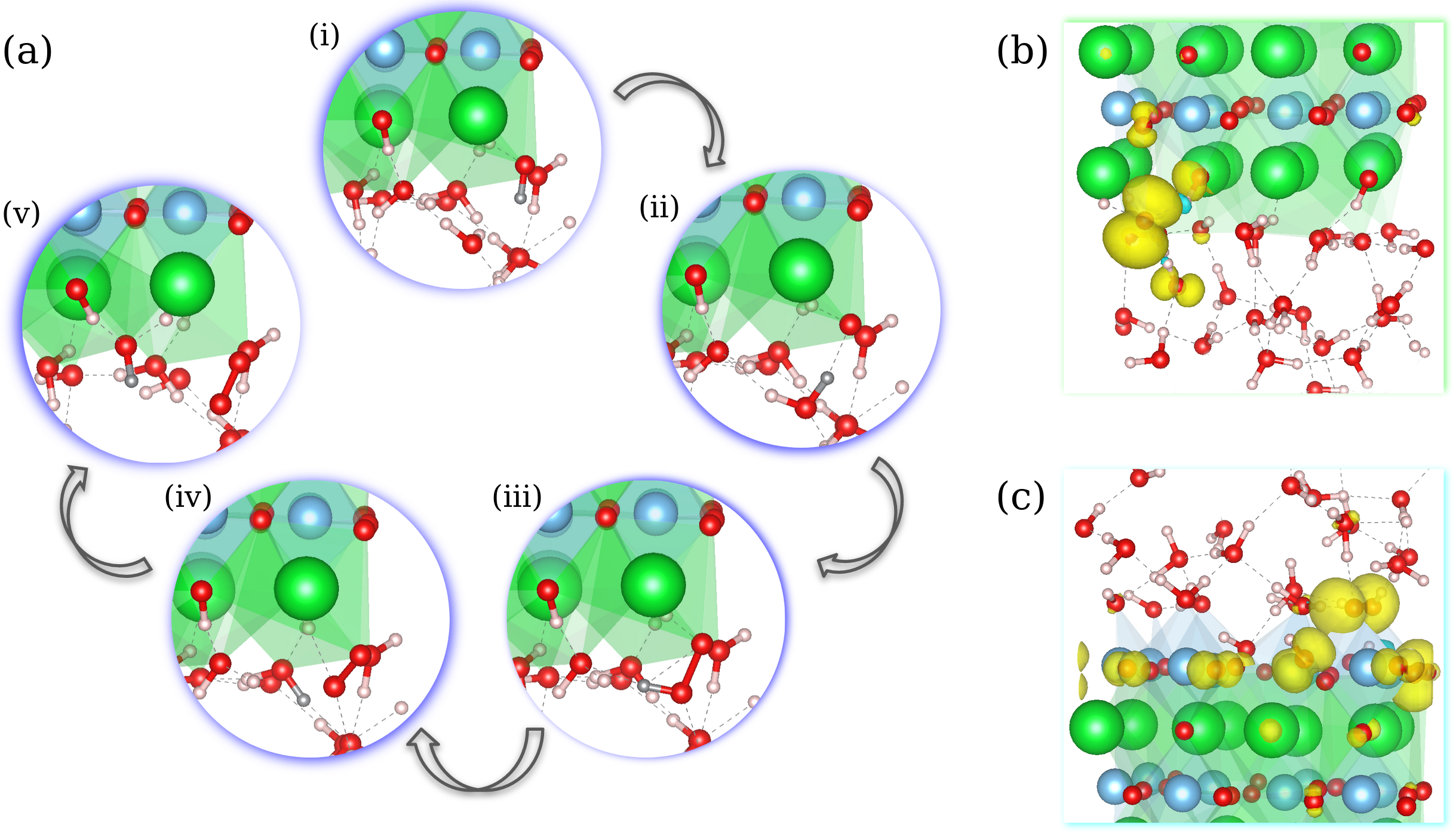}
    \caption{(a) A sequence of four proton-coupled electron transfer (PCET) events reveal the pathway for conversion of (i) OH$^-_{adsorbed}$ to (v) O$_2$ at the SrO-- surface via three reaction intermediates: (ii) Oxygen anion radical (O$^{\bullet-}$), (iii) Hydroperoxyl radical (OOH$^-$), (iv) Superoxide ion (O$_2^{\bullet-}$). The proton removed at each stage of the PCET mechanism is highlighted in grey color. Spin density (in yellow) corresponding to O$^{\bullet-}$ species formed at (b) SrO-terminated and (c) TiO$_2$-terminated aqueous SrTiO$_3$ surfaces.}
    \label{fig:pcet_full}
\end{figure*}

Our proposed four-step PCET cycle for water oxidation at SrTiO$_3$/water interface is shown in Fig. \ref{fig:pcet_full}(a), where (i)-(v) represent relaxed structures obtained after removing a proton coupled with an electron.
The reaction intermediates in the proposed PCET mechanism are identified as: (i) OH$^-$ (adsorbed at a surface Sr/Ti), (ii) O$^{\bullet-}$ (oxygen anion radical), (iii) OOH$^-$ (hydroperoxyl radical), (iv) O$_2^{\bullet-}$ (superoxide ion), and (v) OH$^-$.
Upon removing a proton and an electron from OH$^-$, an oxygen anion radical is formed in the first step, see Fig. \ref{fig:pcet_full}(a, (i)$\longrightarrow$(ii)).
Since there is no longer a proton available at the active species, a neighboring hydrogen-bonded water molecule is chosen and its proton and electron are removed in the second step, (ii)$\longrightarrow$(iii).
A geometric relaxation leads to a spontaneous O--O bond formation resulting in a OOH$^-$ species at the site.
In the third step (iii)$\longrightarrow$(iv), an electron and proton are removed from the intermediate OOH$^-$, which gives rise to a superoxide ion (O$_2^{\bullet-}$) with a shorter O--O bond distance.
As shown in Fig. \ref{fig:pcet_full}(a,iv), the superoxide ion does not immediately dissociate from the surface.
Thus in the final step (iv)$\longrightarrow$(v), a concerted electron-proton transfer from yet another neighboring water molecule results in an OH$^-$.
This newly formed OH$^-$ attacks the active site and replaces the superoxide ion which now leaves its surface position as O$_2$.
%

% \begin{table}[h]
% \centering
% \caption{\label{tab:standardpotentials} Standard one-electron reduction potentials for PCET steps at SrO- and TiO$_2$-terminated aqueous SrTiO$_3$ interface. The values are reported with respect to the normal hydrogen electrode (NHE) scale.}
% %PUT AVG AND STD DEV FOR ERROR IN SRO AND TIO2
% \begin{tabular}{cccccc}
% \hline\hline
%  \multicolumn{1}{c}{Step}&Reaction on&\multicolumn{2}{c}{SrO$-$plane}&\multicolumn{2}{c}{TiO$_2-$plane}\\
%  &SrTiO$_3$-termination&$E^{o}$ (eV)&d$_{OO}$(\AA)&$E^{o}$ (eV)&d$_{OO}$(\AA)\\ \hline
%  1&(i)\mbox{*}OH$^-\xrightarrow[\textcolor{red}{-H^{+}-e^-}]{}$(ii)\mbox{*}O$^{\bullet-}$&$2.157$&(ii) $2.655$&$1.884$&(ii) $2.528$\\
%  2&(ii)\mbox{*}O$^{\bullet-}\xrightarrow[\textcolor{red}{-H^{+}-e^-}]{\textcolor{blue}{+H_2O}}$(iii)\mbox{*}OOH$^-$&$-0.003$&(iii) $1.476$&$1.917$&(iii) $2.236$\\
%  3&(iii)\mbox{*}OOH$^-\xrightarrow[\textcolor{red}{-H^{+}-e^-}]{}$(iv)\mbox{*}O$_2^{\bullet-}$&$0.478$&(iv) $1.350$&$-$&$-$\\
%  4&(iv)\mbox{*}O$_2^{\bullet-}\xrightarrow[\textcolor{red}{-H^{+}-e^-}]{\textcolor{blue}{+H_2O}}$(v)\mbox{*}OH$^-+$ O$_2$&$0.981$&(v) $1.251$&$-$&$-$\\
% \hline\hline
% \end{tabular}
% \end{table}

\begin{table*}
\caption{\label{tab:standardpotentials}Calculated standard one- and two- electron reduction potentials for PCET steps at the SrO- and TiO$_2$-terminated aqueous SrTiO$_3$ interface. The values are reported with respect to the normal hydrogen electrode (NHE) scale. Roman numerals next to each compound refer to the labeling in Fig. \ref{fig:pcet_full}. Entries superscripted with a dagger ($\dagger$) indicate the values for a loosely-bound O$-$OH$^{-}$ as the hydroperoxyl radical is \textit{not} formed in this case.}
\begin{tabular}{cccccccc} \hline\hline
Step & & Reaction on   & \multicolumn{2}{c}{SrO-plane} &  & \multicolumn{2}{c}{TiO$_2$-plane} \\
 & & SrTiO$_3$ termination &  $E^{\circ}$ (eV) & d$_{OO}$ (\AA) &  &  $E^{\circ}$ (eV) & d$_{OO}$ (\AA) \\ \hline
$1$ & & \multicolumn{1}{l}{(ii) $^{*}$O$^{\bullet -}$ \textcolor{red}{+ H$^{+}$ + e$^{-}$} $\to$ (i) $^{*}$OH$^{-}$} & $2.123 \pm 0.028$ & (ii) $2.64 \pm 0.01$ & & $1.85 \pm 0.11$ & (ii) $2.56 \pm 0.04$ \\ 
$2$ & &\multicolumn{1}{l}{(iii) $^{*}$OOH$^{-}$ \textcolor{red}{+ H$^{+}$ + e$^{-}$} $\to$ (ii) $^{*}$O$^{\bullet -}$ $+$ H$_2$O} & $-0.011 \pm 0.008$ & (iii) $1.483 \pm 0.007$ & & $1.93 \pm0.02^\dagger$ & (iii) $2.3 \pm 01^\dagger$ \\ 
$3$ & &\multicolumn{1}{l}{(iv) $^{*}$O$_2^{\bullet-}$ \textcolor{red}{+ H$^{+}$ + e$^{-}$} $\to$ (iii) $^{*}$OOH$^{-}$} & $0.585 \pm 0.107$ & (iv) $1.353 \pm 0.003 $ & & $-$ & $-$ \\ 
$4$ & &\multicolumn{1}{l}{(v) $^{*}$OH$^{-}$ $+$ O$_2$ \textcolor{red}{+ H$^{+}$ + e$^{-}$} $\to$ (iv) $^{*}$O$_2^{\bullet-}$ $+$ H$_2$O} & $0.797 \pm 0.184$ & (v) $1.248 \pm 0.003$ & & $-$ & $-$ \\ \hline
$1 + 2$ & & \multicolumn{1}{l}{$^{*}$OOH$^{-}$ \textcolor{red}{+ $2$H$^{+}$ + $2$e$^{-}$} $\to$ (i) $^{*}$OH$^{-}$ + H$_2$O} & $1.049 \pm 0.007$ & & & $-$ &  \\ 
$3 + 4$ & & \multicolumn{1}{l}{O$_2$ + $^{*}$OH$^{-}$ \textcolor{red}{+ $2$H$^{+}$ + $2$e$^{-}$} $\to$ $^{*}$OOH$^{-}$ + H$_2$O} & $0.67 \pm 0.01$ & & & $-$ & \\ \hline\hline
\end{tabular}	
\end{table*}

We have computed the standard Gibbs free energy changes and the standard one-electron reduction potentials ($E^o$) for two active sites on  SrO- and three on TiO$_2$-terminations, see Supplementary Information Tables S1 and S2.
Table \ref{tab:standardpotentials} shows a comparison of the mean $E^o$ for the water oxidation pathway on the two surfaces.
The mean distance between the interacting oxygens ($d_{OO}$) in any given reaction intermediate is also reported.
On the SrO-terminated surface, the first step which entails the conversion of OH$^-$ to O$^{\bullet-}$ is rate-limiting and requires the maximum reduction potential ($\sim$ 2.1 eV).
In photocatalytic oxidation, the overpotential is the excess energy carried by the hole.
This energy is supplied by the photogenerated holes that carry excess energy supplied from the absorbed photon.
This is often taken as the energy of the valence band edge vs the normal hydrogen electrode (NHE) water oxidation potential at pH $=0$.
As previously shown, our calculations report this energy to be $\sim$ 0.85 eV for a solvated SrTiO$_3$ slab with 50$\%$ TiO$_2$/SrO terminations.
This is not sufficient for the reaction to proceed.
However, the expected overpotential increases with the proportion of TiO$_2$ at the surface.
This explains why in our experiments, all the photocatallytically active samples have a majority of TiO$_2$ composition at the surface. 
The second step on the SrO-surface follows an energetically downhill path, so the formation of OOH$^-$ does not require any energy from the hole.
This also suggests that the two-electron process that takes OH$^-\longrightarrow$ OOH$^-$ directly, could be more favorable than the sequential one-electron processes.
Interestingly, the presence of the intermediate species here proposed has been experimentally observed in these surfaces, albeit without
enough information to identify the two different possible terminations \cite{domingo2019water}.
Table~\ref{tab:standardpotentials} shows that the net potential for a simultaneous two-electron process is $\sim$ 1 eV which makes it energetically accessible for an SrO surface to drive the catalytic cycle with a lower photohole potential.%

If we consider the same PCET-based water oxidation mechanism on the TiO$_2$ surface, the results are rather different.
The first step requires a lower potential than the corresponding step on the SrO side.
However, the consequent step, that is the conversion of O$^{\bullet-}$ to  OOH$^-$ is not supported on the TiO$_2$ surface.
As seen in  Table~\ref{tab:standardpotentials} and in the Supplementary Information, this second PCET reaction
does not result in a O-O bond (the relaxed O-O distance is 2.3 \AA{}).

The reasons for the inadequacy of this surface to catalyze the oxidation of the oxygen anion radical are not obvious.
Figures \ref{fig:pcet_full}(b) and (c) show the spin density of the radical O$^{\bullet-}$ state (Fig. \ref{fig:pcet_full}(a,ii)) on the SrO (b) and TiO$_2$ (c) surfaces.
While this $p_z$-like orbital is highly localized and perpendicular to the  SrO
surface, on TiO$_2$ the orbital is parallel to the surface and much more delocalized within the surface oxygens.
Hence this is a far less reactive species, given that the nearest H$_2$O molecule, needed for the following PCET reaction is oriented along the nodal plane of the O$^{\bullet-}$ radical electron.
%

%
%WAIT FOR SPIN LOCALIZATION RESULTS TO GIVE MORE INSIGHT
%
%However, the difference surface acidity\cite{Wang2012} of the two terminations, manifested in the different observed deprotonation rate would have an effect on the considered reaction. 
%
%The high reduction potential along with the oxygen-oxygen distance for the unformed O--OH$^-$ species are listed in Table \ref{tab:standardpotentials}.
The lack of formation of a crucial intermediate in the PCET mechanism points to an inability of the TiO$_2$ termination to catalyze the oxidation process.
This also explains the lack of reactivity observed on a pure TiO$_2$ termination by AFM experiments, and further supports the observation that an SrO surface is necessary in conjunction with TiO$_2$ for the photocatalytic oxidation of water.

Our theoretical  results clearly indicate that the oxidation reaction proceeds only
in SrO terminated surfaces.
Our experiments also show that Ag$^+$ is reduced at or in the vicinity of these surfaces.
Taken together, these results provide indirect information about the nature and localization of the photoexcited carriers.
They indicate that both the reduction and oxidation reactions happen in close spatial proximity, and hence the exciton does not break before reaching the surface.
This would be consistent with a previously computed exciton radius of $\sim$  5 Bohr radii and a binding energy of 330 eV \cite{PhysRevB.94.041107}.

%as a pure TiO$_2$ termination, by itself, does not support the water oxidation.

%%%%POINTERS BY MARIVI FOR PCET DESCRIPTION%%%%
%chemical reactivity of surfaces; Shen showed GaN using PCET mechanism a full water oxidation mechanism hence use it as a model. different sites; same reaction path; explicit solvent; explain four reactions and how it explains lack of reactivity on tio2 surface; suggests SrO is necessary and pure tio2 does not catalyze.
%trace the same reaction intermediates as Shen et al. as the photocatalytic water-oxidation process remains the same.
%create a smaller table with just SrO & TiO2.

%ADD POINTS OF CONCLUSION FROM THESIS
%ADD A PANEL WITH BAND ALIGNMENT + AVG POTENTIAL OF STO+WATER --> CHECK FIG 9.14 THESIS.

\section{Conclusions}
In conclusion, we have presented a combined experimental and computational study of photocatalytical water splitting activity at SrTiO$_3$ [001] surfaces. 
Using the proxy reaction of Ag$^+$ to Ag reduction we have shown that pure SrO or TiO$_2$ terminations are not photo-active.
Ag is only deposited when mixed terminations are present at the surface and deposition occurs near SrO terminations. 
Our first principles simulations explain these findings and unambiguously show that water oxidation can only occur at SrO surfaces, which efficiently catalyze a four-hole oxidation cycle.  
TiO$_2$ terminations are needed to provide the correct band energy alignment, but they are unfit to catalyze the necessary surface reactions.
These results have important implications for understanding the interplay between surface chemistry and band alignment in semiconductor materials for photocatalytic water splitting.
They highlight the importance of achieving a particular surface nanostructure, ideally through bulk synthesis routes or simple chemical/thermal treatments such as those presented in this work.
We plan to investigate whether this result can be generalized to other perovskite oxide materials, for which ferroelectricity might provide an additional surface alignment and control handle.

\section{Methods}

\subsection{Experiments}

\subsubsection{Water etching procedure}

The etching process takes place within a Parr Microwave Acid Digestion Vessel 4782, a machined PTFE cup sealed within a high-strength, microwave-transparent polymer. The vessel is filled with 20 mL distilled water and the substrate and placed in a Panasonic NN-SN651B inverter microwave oven at power level 3 of 10, for 4 minutes. Etching is followed by annealing in air inside an insulating box on top a MeiVac 2.0” High Temperature Resistive Substrate Heater. As explained in the main text of the manuscript the temperature and duration of the anneals used results in different surfaces.

\subsubsection{Ag deposition}

In our photocatalytic silver deposition experiments, each sample was placed in 0.002 mol AgNO$_3$ solution and illuminated with a 100 W Mercury Vapor lamp at a distance of 60cm for 5 minutes. A UV fused silica ground glass diffuser was placed in front of the sample (Edmund Optics NT49-159).

\subsubsection{Sputtering details}

SrTiO$_{3}$ thin films were grown using off-axis RF Magnetron Sputtering. A 35W power was applied to the 1.3" sputter gun. During the growth an atmosphere of 0.24 mbar with an oxygen to argon ratio of 7/16 was maintained and the sample was kept at an elevated temperature of 560$^{\circ}$C.

\subsection{Computational Methodology}
%\subsection{Simulations}
%\paragraph{DFT simulations and Molecular dynamics.}
\subsubsection{DFT simulations and Molecular dynamics}
%subsubsection{Molecular dynamics}
%VIDUSHI: equilibration for two unit cells and three unit cells (total time), results presented with three unit cells.
The DFT-based \textit{ab initio} molecular dynamics simulations were performed using the \textsc{SIESTA} code with a Generalized Gradient approximate (GGA) exchange-correlation functional.
Specifically, the vdW-BH functional was used which includes dispersive corrections for the van der Waals interactions in the system.
The SrTiO$_3$ slabs have a cross-sectional area of $2\sqrt{2}a_0 \times 2\sqrt{2}a_0$ where the (bulk) lattice constant is $a_0=3.909$ \AA.

A double-zeta polarized (DZP) basis set was used for the electronic wavefunctions of Sr and Ti, while the O orbitals were described using a long-range quadruple-zeta basis set (QZP).
The size of the oxygen basis set was found to have a significant effect on the band-edge positions as the valence band edge mostly consists of O$-2p$ bands.
All calculations were spin-polarized, and the GGA+U computations included a Hubbard U correction term of $4.45$ eV for the Ti$-3d$ states \cite{Marzari2020}.
The $(001)$ SrTiO$_3$ surface calculations were performed with 13-layer surface slabs for both the vacuum-interfacing symmetric SrO- and TiO$_2$- terminated structures, as shown in supplemental figure, Fig. S5.
A vacuum region of 15 \AA\, (along the $z-$axis) was found to be sufficient to separate the two nonpolar surfaces of a symmetric SrTiO$_3$ slab.
The structures used for determining the band edge positions in Fig. \ref{fig:bandalignment} were geometrically relaxed until the remnant forces in the system were less than $0.01$ eV/\AA.
While a large set of simulations in this work were performed with a computationally-efficient GGA-type exchange-correlation functional, the sensitivity of band-related properties to the hierarchy of functional \cite{Car2016} used is widely recognized \cite{Ma2020, Kharche2014}.
Hence, we estimate the band-edge positions using a range-separated hybrid functional, \textsc{HSE06} \cite{HSE1, HSE2}, as implemented in \textsc{VASP} \cite{Kresse1996}.
A plane-wave cutoff of 500 eV was used with the projector augmented-wave (\textsc{PAW}) method of potentials \cite{PAW} in \textsc{VASP} and the reciprocal space was sampled with a single $k$ point at $\Gamma$.

The interfacial properties and structure of aqueous SrTiO$_3$ surfaces were analyzed with the help of DFT-based molecular dynamics (MD) simulations.
The system consisted of a $[2\sqrt{2}\times2\sqrt{2}\times3]$ nonpolar $(001)$ SrTiO$_3$ slab with a box of 64 H$_2$O molecules adjacently placed along the $z-$axis.
Unlike the case of SrTiO$_3$/vacuum symmetric interfaces, in the fully solvated SrTiO$_3$/H$_2$O system, water makes contact with two different chemical terminations (SrO/H$_2$O and TiO$_2$/H$_2$O) at once, see Fig. \ref{fig:sto_waterfull}(a).
In our MD simulations of solvated SrTiO$_3$ system, the SrTiO$_3$ slab comprised 3-layers each of SrO and TiO$_2$.
%We have also performed similar MD simulations with 2-layers each of SrO and TiO$_2$, and... 
The SrO- and TiO$_2$- terminations were separated by $\sim18$ \AA\, region of water.
Based on the (planar-averaged) macroscopic electrostatic potential of the solvated system, it was confirmed that the water region screens the two asymmetric-terminated surfaces and that there is no net charge transfer from one surface to the other, see supplemental figure Fig. S3.

At the beginning of MD simulation, the surfaces were non-hydroxylated (no water dissociation).
A GGA-type vdW-BH functional was used in \textsc{SIESTA} to perform a 20 picoseconds (ps) long MD simulation with a time-step of 0.5 femtoseconds (fs).
During this MD, the hydrated slab is annealed to a temperature of $T=330$ K using a velocity rescaling thermostat, and water interacts with TiO$_2$ surface on one side and SrO on the other giving different amounts of dissociation on both sides.

%PCET analysis picked up structures from equilibrated MD.

\subsubsection{Band Alignment}
%The subject of energy level alignment for SrTiO$_3$ surfaces using DFT methods has been the focus of many recent studies \cite{Ma2020,Marzari2020,Kim2015}.
%
%It was concluded that the commonly used GGA-type functionals such as PBE, result in an inaccurate prediction of the work functions for SrO- and TiO$_2$-terminated surfaces.
%
%Relative differences between the the two terminations are predicted within less than 0.2 eV error by PBE \cite{Ma2020}.
%
%On the other hand, a range-separated hybrid functional such as HSE06 \cite{HSE1, HSE2} is more accurate in predicting the energy levels albeit computationally expensive.
%
%Many times a good trade-off between accuracy and computational cost is attained with the use of Hubbard correction to the DFT energies (DFT+U) \cite{Marzari2020, Kim2015}.
%Recent studies \cite{Marzari2020, Kim2015} have also shown that band energy levels computed using DFT+U are also much more accurate.
%
Our current work focuses on a qualitative understanding of the energy level alignment for different (001) SrTiO$_3$ surface terminations in conjunction with vacuum and water.
Hence, several components of this study employ the GGA and GGA+U levels of theory, where we choose $U=4.45$ eV following the work of ref\cite{Marzari2020}.
In order to further refine the results of our GGA-based simulations, we obtain the band edges of the SrTiO$_3$ slab systems with HSE06 in Fig. \ref{fig:bandalignment}.
This allows for a more accurate quantitative comparison of the energy levels.
Extended results obtained with DFT+U are presented in the Supplementary information, see Fig. S4. 
These results also include level alignments for slabs solvated with a single monolayer of water.

The band edges for SrTiO$_3$/vac. slabs are determined using a surface-vacuum alignment technique in which the VBE is given by the difference between the vacuum level derived from the electrostatic potential of the slab model and the highest occupied level (HOMO) in the system.
Similarly, the conduction band edge is given by the difference between the vacuum level and the lowest unoccupied energy level (LUMO) of the slab.
The band alignments for the symmetric terminations of SrO/vac and TiO$_2$/vac are given in Supplemental section S2.

For the fully solvated SrTiO$_3$/H$_2$O interfaces, 
%the band edge positions are well-defined with respect to vacuum on the semiconductor side in a manner outlined above.
we obtain the position of the $1b_1$ peak, which marks the highest occupied state in the bulk-region of water and reference it to the vacuum obtained from a pure water-slab calculation.
This amounts to an energy-shift of the kind: $(E_{1b_1, bulk})_{\textrm{STO/Water}} \longrightarrow (E_{1b_1})_{\textrm{Water-slab}}$.
The band edge energy (HOMO and LUMO) levels of the full SrTiO$_3$/H$_2$O system are then aligned with respect to the $1b_1$ level of water.
In the absence of an exact `vacuum level' in the case of solvated SrTiO$_3$ surfaces, the $1b_1$ level serves as a natural datum against which the band energies can be compared.
A detailed account of the band alignment in SrTiO$_3$/water surfaces is given in the Supplemental section S4. 

\subsubsection{Modelling of photocatalytic water oxidation}
We consider several SrO- and TiO$_2$-terminated water-oxidation sites and the associated redox potentials are reported in Supplemental Tables S1, S2.
In our \textit{ab initio} simulations, we do not explicitly model a photon adsorption event or the photogenerated charge carrier separation.
We assume that the hole while being generated in the bulk of the catalyst (SrTiO$_3$) becomes available at the top of the valence band to oxidize water and release oxygen.
At each step of our proposed water oxidation mechanism, one is given to understand that an electron is removed from the active site region (in our case, the SrTiO$_3$ slab) filling a photohole generated in the aqueous reservoir.
While the charge transfer phenomenon at the intersection of SrTiO$_3$ slab and physisorbed water is of primary interest, we find that the interaction between surface-dissociated water species and the remainder bulk solvent plays a crucial mechanistic role too.
Hence in this study we go beyond the implicit-solvent model of Shen {\it et al.} \cite{Shen2010} and explicitly consider all of the water molecules in the solvated SrTiO$_3$ system.
The presence of bulk water beyond the first adsorbed monolayer at the catalyst surface acts as a reservoir for coupling the loss of a proton with an oxidation event.
From a single equilibrated MD trajectory, we select a snapshot such as Fig. \ref{fig:pcet_full}(i), and evaluate the completion of the proposed cycle.
Following this, at each PCET step, a proton and an electron (that is, a hydrogen atom) are removed from either a hydroxide ion dissociated on the SrTiO$_3$ surface or a water molecule that is hydrogen-bonded to the target surface intermediate.
As the geometric optimization of each intermediate structure is carried out at $T=0$ K, the temperature effects are not considered explicitly, however, the initial structure of a PCET cycle is always derived from a thermostated parent trajectory.
Apart from considering a fully solvated periodic system, 
our calculations of free energy changes and standard reduction potentials follow a similar procedure as outlined in our previous work by Shen {\it et al.} \cite{Shen2010}.
%
%We note that, at each reaction step, we follow the same procedure described in Shen {\it et al.}\cite{Shen2010}.
%
In addition, we allow for a complete relaxation of the solvent in the system, and after each proton removal we observe that the water molecules near the active site rearrange to initiate the following reaction in the cycle.
We also note that, despite using a semi-local exchange and correlation (XC) potential, namely vdW-BH \cite{BH1989}, the localization of the photohole on the initial reaction species is rather independent of the XC potential \cite{Sharma2020}. We also confirmed that systems (i) and (iii) in Fig. \ref{fig:pcet_full} are singlets and systems (ii) and (iv) are electronic doublets. 

\paragraph{Gibbs free energy and Standard reduction potentials.}
The standard redox potentials were computed with respect to the normal hydrogen electrode (NHE).
As a photocatalyst, SrTiO$_3$ supplies an overpotential defined in this case as the difference between VBE and NHE \cite{Takanabe2017, Li2015}.
The standard free energy for the one-electron hydrogen oxidation reaction at NHE is given by,
\begin{align}\begin{split}
    \Delta G_{NHE} &= G_s(H^+) + G_g(e^-) - \frac{G_g(H_2)}{2} ~,\label{eq:pcetone}\\
    G_s(H^+) &= G_g(H^+) + \Delta G_{solv.}(H^+) ~,
    \end{split}
\end{align}
where $G_g(H_2) = -31.386$ eV, is the standard free energy of a gas-phase H$_2$ molecule computed using methods consistent with the other calculations in this work.
$G_g(H_2)$ also contains the contribution of vibrational zero-point energies (ZPEs) \cite{NIST}.
The standard free energy of a gas-phase electron, $G_g(e^-)=-0.0376$ eV, was obtained using the Fermi-Dirac statistics \cite{Bartmess1994}.
The standard free energy of a solvated proton, $G_s(H^+)=-11.719$ eV, was computed using the free energy of a gas-phase proton $G_g(H^+)$, and the experimental solvation energy of the proton $\Delta G_{solv.}(H^+)$ \cite{McQuarrie1975, Tissandier1998}.
Thus, using Eq.~\eqref{eq:pcetone} we obtain the standard oxidation potential of the NHE, $E^o_{NHE}=-\Delta G_{NHE}=-3.936$ eV.
The Gibbs free energy change computed at each step of the PCET cycle ($\Delta G_I$) include the ZPEs of the respective reaction intermediates. On the absolute scale,
\begin{align}\begin{split}
	\Delta G_{abs,1} &= G(ii) - G(i) - E_{zpe}(OH^-) \\
	                 & + G_s(H^+) + G_g(e^-) ~,\\
	\Delta G_{abs,2} &= G(iii) - G(ii) + E_{zpe}(OOH^-) - E_{zpe}(H_2O) \\ &\quad + G_s(H^+) + G_g(e^-) ~,\\
	\Delta G_{abs,3} &= G(iv) - G(iii) + E_{zpe}(O_2^{\bullet-}) - E_{zpe}(OOH^-) \\ &\quad + G_s(H^+) + G_g(e^-) ~,\\
	\Delta G_{abs,4} &= G(v) - G(iv) + E_{zpe}(OH^-) + E_{zpe}(O_2) \\ &\quad  - E_{zpe}(O_2^{\bullet-}) - E_{zpe}(H_2O) \\
	                 & + G_s(H^+) + G_g(e^-) ~,
	\end{split}
\end{align}
where the successive PCET steps differ in one proton and electron ($H^+ + e^-$) pair, and the second and fourth steps borrown a H$_2$O molecule from the interacting bulk water reservoir.
On the physical (NHE) scale, the Gibbs free energy changes and the standard one-electron reduction potentials translate to $\Delta G_I = E^o = \Delta G_{abs, I} - \Delta G_{NHE}$.
A detailed description of the water-oxidation PCET mechanism and the band alignment at the interface of SrTiO$_3$/water can be found in the Supplementary Information and in \cite{vidushithesis}.

Input files and Jupyter notebooks outlines various calculations are provided in a data repository associated with this manuscript \cite{data}.

\begin{acknowledgments}
M.D, A.L and B.B were supported by NSF DMR-1055413 and DMR-1334867.
M. F-S and V.S were funded by  the U.S. Department of Energy, Office of Science, Basic Energy Sciences, under Awards DE-SC0001137 and DE-SC0019394, as part of the CCS and CTC Programs.
The authors thank Stony Brook Research Computing and Cyberinfrastructure, and the Institute for Advanced Computational Science at Stony Brook University for access to the high-performance SeaWulf computing system, which was made possible by a \$1.4M National Science Foundation grant (\#1531492). B.P. acknowledges National Science Foundation (Platform for the Accelerated Realization, Analysis, and Discovery of Interface Materials (PARADIM)) under Cooperative Agreement No. DMR-2039380. 
CED acknowledges support from the National Science Foundation under Grant No. DMR-1918455. The Flatiron Institute is a division of the Simons Foundation.
\end{acknowledgments}

\pagebreak

\bibliography{references}

\begin{thebibliography}{68}%
\makeatletter
\providecommand \@ifxundefined [1]{%
 \@ifx{#1\undefined}
}%
\providecommand \@ifnum [1]{%
 \ifnum #1\expandafter \@firstoftwo
 \else \expandafter \@secondoftwo
 \fi
}%
\providecommand \@ifx [1]{%
 \ifx #1\expandafter \@firstoftwo
 \else \expandafter \@secondoftwo
 \fi
}%
\providecommand \natexlab [1]{#1}%
\providecommand \enquote  [1]{``#1''}%
\providecommand \bibnamefont  [1]{#1}%
\providecommand \bibfnamefont [1]{#1}%
\providecommand \citenamefont [1]{#1}%
\providecommand \href@noop [0]{\@secondoftwo}%
\providecommand \href [0]{\begingroup \@sanitize@url \@href}%
\providecommand \@href[1]{\@@startlink{#1}\@@href}%
\providecommand \@@href[1]{\endgroup#1\@@endlink}%
\providecommand \@sanitize@url [0]{\catcode `\\12\catcode `\$12\catcode
  `\&12\catcode `\#12\catcode `\^12\catcode `\_12\catcode `\%12\relax}%
\providecommand \@@startlink[1]{}%
\providecommand \@@endlink[0]{}%
\providecommand \url  [0]{\begingroup\@sanitize@url \@url }%
\providecommand \@url [1]{\endgroup\@href {#1}{\urlprefix }}%
\providecommand \urlprefix  [0]{URL }%
\providecommand \Eprint [0]{\href }%
\providecommand \doibase [0]{https://doi.org/}%
\providecommand \selectlanguage [0]{\@gobble}%
\providecommand \bibinfo  [0]{\@secondoftwo}%
\providecommand \bibfield  [0]{\@secondoftwo}%
\providecommand \translation [1]{[#1]}%
\providecommand \BibitemOpen [0]{}%
\providecommand \bibitemStop [0]{}%
\providecommand \bibitemNoStop [0]{.\EOS\space}%
\providecommand \EOS [0]{\spacefactor3000\relax}%
\providecommand \BibitemShut  [1]{\csname bibitem#1\endcsname}%
\let\auto@bib@innerbib\@empty
%</preamble>
\bibitem [{\citenamefont {Kudo}\ and\ \citenamefont {Miseki}(2009)}]{Kudo2009}%
  \BibitemOpen
  \bibfield  {author} {\bibinfo {author} {\bibfnamefont {A.}~\bibnamefont
  {Kudo}}\ and\ \bibinfo {author} {\bibfnamefont {Y.}~\bibnamefont {Miseki}},\
  }\href {https://doi.org/10.1039/B800489G} {\bibfield  {journal} {\bibinfo
  {journal} {Chem. Soc. Rev.}\ }\textbf {\bibinfo {volume} {38}},\ \bibinfo
  {pages} {253} (\bibinfo {year} {2009})}\BibitemShut {NoStop}%
\bibitem [{\citenamefont {Maeda}\ and\ \citenamefont
  {Domen}(2007)}]{Maeda2007}%
  \BibitemOpen
  \bibfield  {author} {\bibinfo {author} {\bibfnamefont {K.}~\bibnamefont
  {Maeda}}\ and\ \bibinfo {author} {\bibfnamefont {K.}~\bibnamefont {Domen}},\
  }\href {https://doi.org/10.1021/jp070911w} {\bibfield  {journal} {\bibinfo
  {journal} {The Journal of Physical Chemistry C}\ }\textbf {\bibinfo {volume}
  {111}},\ \bibinfo {pages} {7851} (\bibinfo {year} {2007})}\BibitemShut
  {NoStop}%
\bibitem [{\citenamefont {Rousseau}\ \emph {et~al.}(2020)\citenamefont
  {Rousseau}, \citenamefont {Glezakou},\ and\ \citenamefont
  {Selloni}}]{Selloni2020}%
  \BibitemOpen
  \bibfield  {author} {\bibinfo {author} {\bibfnamefont {R.}~\bibnamefont
  {Rousseau}}, \bibinfo {author} {\bibfnamefont {V.-A.}\ \bibnamefont
  {Glezakou}},\ and\ \bibinfo {author} {\bibfnamefont {A.}~\bibnamefont
  {Selloni}},\ }\href {https://doi.org/10.1038/s41578-020-0198-9} {\bibfield
  {journal} {\bibinfo  {journal} {Nature Reviews Materials}\ }\textbf {\bibinfo
  {volume} {5}},\ \bibinfo {pages} {460} (\bibinfo {year} {2020})}\BibitemShut
  {NoStop}%
\bibitem [{\citenamefont {Linsebigler}\ \emph {et~al.}(1995)\citenamefont
  {Linsebigler}, \citenamefont {Lu},\ and\ \citenamefont
  {Yates}}]{Linsebigler1995}%
  \BibitemOpen
  \bibfield  {author} {\bibinfo {author} {\bibfnamefont {A.~L.}\ \bibnamefont
  {Linsebigler}}, \bibinfo {author} {\bibfnamefont {G.}~\bibnamefont {Lu}},\
  and\ \bibinfo {author} {\bibfnamefont {J.~T.}\ \bibnamefont {Yates}},\ }\href
  {https://doi.org/10.1021/cr00035a013} {\bibfield  {journal} {\bibinfo
  {journal} {Chemical Reviews}\ }\textbf {\bibinfo {volume} {95}},\ \bibinfo
  {pages} {735} (\bibinfo {year} {1995})}\BibitemShut {NoStop}%
\bibitem [{\citenamefont {Eidsv{\aa}g}\ \emph {et~al.}(2021)\citenamefont
  {Eidsv{\aa}g}, \citenamefont {Bentouba}, \citenamefont {Vajeeston},
  \citenamefont {Yohi},\ and\ \citenamefont {Velauthapillai}}]{Eidsvag2021}%
  \BibitemOpen
  \bibfield  {author} {\bibinfo {author} {\bibfnamefont {H.}~\bibnamefont
  {Eidsv{\aa}g}}, \bibinfo {author} {\bibfnamefont {S.}~\bibnamefont
  {Bentouba}}, \bibinfo {author} {\bibfnamefont {P.}~\bibnamefont {Vajeeston}},
  \bibinfo {author} {\bibfnamefont {S.}~\bibnamefont {Yohi}},\ and\ \bibinfo
  {author} {\bibfnamefont {D.}~\bibnamefont {Velauthapillai}},\ }\bibfield
  {journal} {\bibinfo  {journal} {Molecules}\ }\textbf {\bibinfo {volume}
  {26}},\ \href {https://doi.org/10.3390/molecules26061687}
  {10.3390/molecules26061687} (\bibinfo {year} {2021})\BibitemShut {NoStop}%
\bibitem [{\citenamefont {Miyoshi}\ \emph {et~al.}(2018)\citenamefont
  {Miyoshi}, \citenamefont {Nishioka},\ and\ \citenamefont
  {Maeda}}]{Miyoshi2018}%
  \BibitemOpen
  \bibfield  {author} {\bibinfo {author} {\bibfnamefont {A.}~\bibnamefont
  {Miyoshi}}, \bibinfo {author} {\bibfnamefont {S.}~\bibnamefont {Nishioka}},\
  and\ \bibinfo {author} {\bibfnamefont {K.}~\bibnamefont {Maeda}},\ }\href
  {https://doi.org/https://doi.org/10.1002/chem.201800799} {\bibfield
  {journal} {\bibinfo  {journal} {Chemistry – A European Journal}\ }\textbf
  {\bibinfo {volume} {24}},\ \bibinfo {pages} {18204} (\bibinfo {year}
  {2018})}\BibitemShut {NoStop}%
\bibitem [{\citenamefont {Rajeshwar}(2007)}]{Rajeshwar2007}%
  \BibitemOpen
  \bibfield  {author} {\bibinfo {author} {\bibfnamefont {K.}~\bibnamefont
  {Rajeshwar}},\ }\href {https://doi.org/10.1007/s10800-007-9333-1} {\bibfield
  {journal} {\bibinfo  {journal} {Journal of Applied Electrochemistry}\
  }\textbf {\bibinfo {volume} {37}},\ \bibinfo {pages} {765} (\bibinfo {year}
  {2007})}\BibitemShut {NoStop}%
\bibitem [{\citenamefont {Lee}\ \emph {et~al.}(2021)\citenamefont {Lee},
  \citenamefont {Wang}, \citenamefont {Zhou}, \citenamefont {Tong},
  \citenamefont {Liu}, \citenamefont {Galli},\ and\ \citenamefont
  {Choi}}]{Lee2021}%
  \BibitemOpen
  \bibfield  {author} {\bibinfo {author} {\bibfnamefont {D.}~\bibnamefont
  {Lee}}, \bibinfo {author} {\bibfnamefont {W.}~\bibnamefont {Wang}}, \bibinfo
  {author} {\bibfnamefont {C.}~\bibnamefont {Zhou}}, \bibinfo {author}
  {\bibfnamefont {X.}~\bibnamefont {Tong}}, \bibinfo {author} {\bibfnamefont
  {M.}~\bibnamefont {Liu}}, \bibinfo {author} {\bibfnamefont {G.}~\bibnamefont
  {Galli}},\ and\ \bibinfo {author} {\bibfnamefont {K.-S.}\ \bibnamefont
  {Choi}},\ }\href {https://doi.org/10.1038/s41560-021-00777-x} {\bibfield
  {journal} {\bibinfo  {journal} {Nature Energy}\ }\textbf {\bibinfo {volume}
  {6}},\ \bibinfo {pages} {287} (\bibinfo {year} {2021})}\BibitemShut {NoStop}%
\bibitem [{\citenamefont {Kanhere}\ and\ \citenamefont
  {Chen}(2014)}]{Kanhere2014}%
  \BibitemOpen
  \bibfield  {author} {\bibinfo {author} {\bibfnamefont {P.}~\bibnamefont
  {Kanhere}}\ and\ \bibinfo {author} {\bibfnamefont {Z.}~\bibnamefont {Chen}},\
  }\href {https://doi.org/10.3390/molecules191219995} {\bibfield  {journal}
  {\bibinfo  {journal} {Molecules}\ }\textbf {\bibinfo {volume} {19}},\
  \bibinfo {pages} {19995} (\bibinfo {year} {2014})}\BibitemShut {NoStop}%
\bibitem [{\citenamefont {Li}\ \emph {et~al.}(2013)\citenamefont {Li},
  \citenamefont {Zhang}, \citenamefont {Wang}, \citenamefont {Yang},
  \citenamefont {Li}, \citenamefont {Zhu}, \citenamefont {Zhou}, \citenamefont
  {Han},\ and\ \citenamefont {Li}}]{Li2013}%
  \BibitemOpen
  \bibfield  {author} {\bibinfo {author} {\bibfnamefont {R.}~\bibnamefont
  {Li}}, \bibinfo {author} {\bibfnamefont {F.}~\bibnamefont {Zhang}}, \bibinfo
  {author} {\bibfnamefont {D.}~\bibnamefont {Wang}}, \bibinfo {author}
  {\bibfnamefont {J.}~\bibnamefont {Yang}}, \bibinfo {author} {\bibfnamefont
  {M.}~\bibnamefont {Li}}, \bibinfo {author} {\bibfnamefont {J.}~\bibnamefont
  {Zhu}}, \bibinfo {author} {\bibfnamefont {X.}~\bibnamefont {Zhou}}, \bibinfo
  {author} {\bibfnamefont {H.}~\bibnamefont {Han}},\ and\ \bibinfo {author}
  {\bibfnamefont {C.}~\bibnamefont {Li}},\ }\href
  {https://doi.org/10.1038/ncomms2401} {\bibfield  {journal} {\bibinfo
  {journal} {Nature Communications}\ }\textbf {\bibinfo {volume} {4}},\
  \bibinfo {pages} {1432} (\bibinfo {year} {2013})}\BibitemShut {NoStop}%
\bibitem [{\citenamefont {Wrighton}\ \emph {et~al.}(1976)\citenamefont
  {Wrighton}, \citenamefont {Ellis}, \citenamefont {Wolczanski}, \citenamefont
  {Morse}, \citenamefont {Abrahamson},\ and\ \citenamefont
  {Ginley}}]{Wrighton1976}%
  \BibitemOpen
  \bibfield  {author} {\bibinfo {author} {\bibfnamefont {M.~S.}\ \bibnamefont
  {Wrighton}}, \bibinfo {author} {\bibfnamefont {A.~B.}\ \bibnamefont {Ellis}},
  \bibinfo {author} {\bibfnamefont {P.~T.}\ \bibnamefont {Wolczanski}},
  \bibinfo {author} {\bibfnamefont {D.~L.}\ \bibnamefont {Morse}}, \bibinfo
  {author} {\bibfnamefont {H.~B.}\ \bibnamefont {Abrahamson}},\ and\ \bibinfo
  {author} {\bibfnamefont {D.~S.}\ \bibnamefont {Ginley}},\ }\href
  {https://doi.org/10.1021/ja00426a017} {\bibfield  {journal} {\bibinfo
  {journal} {Journal of the American Chemical Society}\ }\textbf {\bibinfo
  {volume} {98}},\ \bibinfo {pages} {2774} (\bibinfo {year}
  {1976})}\BibitemShut {NoStop}%
\bibitem [{\citenamefont {van Benthem}\ \emph {et~al.}(2001)\citenamefont {van
  Benthem}, \citenamefont {Els{\"a}sser},\ and\ \citenamefont
  {French}}]{vanBenthem2001}%
  \BibitemOpen
  \bibfield  {author} {\bibinfo {author} {\bibfnamefont {K.}~\bibnamefont {van
  Benthem}}, \bibinfo {author} {\bibfnamefont {C.}~\bibnamefont
  {Els{\"a}sser}},\ and\ \bibinfo {author} {\bibfnamefont {R.~H.}\ \bibnamefont
  {French}},\ }\href {https://doi.org/10.1063/1.1415766} {\bibfield  {journal}
  {\bibinfo  {journal} {Journal of Applied Physics}\ }\textbf {\bibinfo
  {volume} {90}},\ \bibinfo {pages} {6156} (\bibinfo {year}
  {2001})}\BibitemShut {NoStop}%
\bibitem [{\citenamefont {Giocondi}\ and\ \citenamefont
  {Rohrer}(2003)}]{Giocondi2003}%
  \BibitemOpen
  \bibfield  {author} {\bibinfo {author} {\bibfnamefont {J.~L.}\ \bibnamefont
  {Giocondi}}\ and\ \bibinfo {author} {\bibfnamefont {G.~S.}\ \bibnamefont
  {Rohrer}},\ }\href {https://doi.org/10.1111/j.1151-2916.2003.tb03445.x}
  {\bibfield  {journal} {\bibinfo  {journal} {Journal of the American Ceramic
  Society}\ }\textbf {\bibinfo {volume} {86}},\ \bibinfo {pages} {1182}
  (\bibinfo {year} {2003})}\BibitemShut {NoStop}%
\bibitem [{\citenamefont {Kato}\ \emph {et~al.}(2013)\citenamefont {Kato},
  \citenamefont {Kobayashi}, \citenamefont {Hara},\ and\ \citenamefont
  {Kakihana}}]{Kato2013}%
  \BibitemOpen
  \bibfield  {author} {\bibinfo {author} {\bibfnamefont {H.}~\bibnamefont
  {Kato}}, \bibinfo {author} {\bibfnamefont {M.}~\bibnamefont {Kobayashi}},
  \bibinfo {author} {\bibfnamefont {M.}~\bibnamefont {Hara}},\ and\ \bibinfo
  {author} {\bibfnamefont {M.}~\bibnamefont {Kakihana}},\ }\href
  {https://doi.org/10.1039/C3CY00014A} {\bibfield  {journal} {\bibinfo
  {journal} {Catal. Sci. Technol.}\ }\textbf {\bibinfo {volume} {3}},\ \bibinfo
  {pages} {1733} (\bibinfo {year} {2013})}\BibitemShut {NoStop}%
\bibitem [{\citenamefont {Kato}\ and\ \citenamefont {Kudo}(2002)}]{Kato2002}%
  \BibitemOpen
  \bibfield  {author} {\bibinfo {author} {\bibfnamefont {H.}~\bibnamefont
  {Kato}}\ and\ \bibinfo {author} {\bibfnamefont {A.}~\bibnamefont {Kudo}},\
  }\href {https://doi.org/10.1021/jp0255482} {\bibfield  {journal} {\bibinfo
  {journal} {The Journal of Physical Chemistry B}\ }\textbf {\bibinfo {volume}
  {106}},\ \bibinfo {pages} {5029} (\bibinfo {year} {2002})}\BibitemShut
  {NoStop}%
\bibitem [{\citenamefont {Wang}\ \emph {et~al.}(2006)\citenamefont {Wang},
  \citenamefont {Ye}, \citenamefont {Kako},\ and\ \citenamefont
  {Kimura}}]{Wang2006}%
  \BibitemOpen
  \bibfield  {author} {\bibinfo {author} {\bibfnamefont {D.}~\bibnamefont
  {Wang}}, \bibinfo {author} {\bibfnamefont {J.}~\bibnamefont {Ye}}, \bibinfo
  {author} {\bibfnamefont {T.}~\bibnamefont {Kako}},\ and\ \bibinfo {author}
  {\bibfnamefont {T.}~\bibnamefont {Kimura}},\ }\href
  {https://doi.org/10.1021/jp062487p} {\bibfield  {journal} {\bibinfo
  {journal} {The Journal of Physical Chemistry B}\ }\textbf {\bibinfo {volume}
  {110}},\ \bibinfo {pages} {15824} (\bibinfo {year} {2006})}\BibitemShut
  {NoStop}%
\bibitem [{\citenamefont {Wei}\ \emph {et~al.}(2009)\citenamefont {Wei},
  \citenamefont {Dai}, \citenamefont {Jin},\ and\ \citenamefont
  {Huang}}]{Wei2009}%
  \BibitemOpen
  \bibfield  {author} {\bibinfo {author} {\bibfnamefont {W.}~\bibnamefont
  {Wei}}, \bibinfo {author} {\bibfnamefont {Y.}~\bibnamefont {Dai}}, \bibinfo
  {author} {\bibfnamefont {H.}~\bibnamefont {Jin}},\ and\ \bibinfo {author}
  {\bibfnamefont {B.}~\bibnamefont {Huang}},\ }\href
  {https://doi.org/10.1088/0022-3727/42/5/055401} {\bibfield  {journal}
  {\bibinfo  {journal} {Journal of Physics D: Applied Physics}\ }\textbf
  {\bibinfo {volume} {42}},\ \bibinfo {pages} {055401} (\bibinfo {year}
  {2009})}\BibitemShut {NoStop}%
\bibitem [{\citenamefont {Niishiro}\ \emph {et~al.}(2005)\citenamefont
  {Niishiro}, \citenamefont {Kato},\ and\ \citenamefont {Kudo}}]{Niishiro2005}%
  \BibitemOpen
  \bibfield  {author} {\bibinfo {author} {\bibfnamefont {R.}~\bibnamefont
  {Niishiro}}, \bibinfo {author} {\bibfnamefont {H.}~\bibnamefont {Kato}},\
  and\ \bibinfo {author} {\bibfnamefont {A.}~\bibnamefont {Kudo}},\ }\href
  {https://doi.org/10.1039/B502147B} {\bibfield  {journal} {\bibinfo  {journal}
  {Phys. Chem. Chem. Phys.}\ }\textbf {\bibinfo {volume} {7}},\ \bibinfo
  {pages} {2241} (\bibinfo {year} {2005})}\BibitemShut {NoStop}%
\bibitem [{\citenamefont {Konta}\ \emph {et~al.}(2004)\citenamefont {Konta},
  \citenamefont {Ishii}, \citenamefont {Kato},\ and\ \citenamefont
  {Kudo}}]{Konta2004}%
  \BibitemOpen
  \bibfield  {author} {\bibinfo {author} {\bibfnamefont {R.}~\bibnamefont
  {Konta}}, \bibinfo {author} {\bibfnamefont {T.}~\bibnamefont {Ishii}},
  \bibinfo {author} {\bibfnamefont {H.}~\bibnamefont {Kato}},\ and\ \bibinfo
  {author} {\bibfnamefont {A.}~\bibnamefont {Kudo}},\ }\href
  {https://doi.org/10.1021/jp049556p} {\bibfield  {journal} {\bibinfo
  {journal} {The Journal of Physical Chemistry B}\ }\textbf {\bibinfo {volume}
  {108}},\ \bibinfo {pages} {8992} (\bibinfo {year} {2004})}\BibitemShut
  {NoStop}%
\bibitem [{\citenamefont {Ohsawa}\ \emph {et~al.}(2006)\citenamefont {Ohsawa},
  \citenamefont {Nakajima}, \citenamefont {Matsumoto},\ and\ \citenamefont
  {Koinuma}}]{Ohsawa2006}%
  \BibitemOpen
  \bibfield  {author} {\bibinfo {author} {\bibfnamefont {T.}~\bibnamefont
  {Ohsawa}}, \bibinfo {author} {\bibfnamefont {K.}~\bibnamefont {Nakajima}},
  \bibinfo {author} {\bibfnamefont {Y.}~\bibnamefont {Matsumoto}},\ and\
  \bibinfo {author} {\bibfnamefont {H.}~\bibnamefont {Koinuma}},\ }\href
  {https://doi.org/https://doi.org/10.1016/j.apsusc.2005.05.086} {\bibfield
  {journal} {\bibinfo  {journal} {Applied Surface Science}\ }\textbf {\bibinfo
  {volume} {252}},\ \bibinfo {pages} {2603} (\bibinfo {year} {2006})},\
  \bibinfo {note} {proceedings of the Third Japan-US Workshop on Combinatorial
  Material Science and Technology}\BibitemShut {NoStop}%
\bibitem [{\citenamefont {Tanaka}\ \emph {et~al.}(2010)\citenamefont {Tanaka},
  \citenamefont {Takata}, \citenamefont {Katayama}, \citenamefont {Takahashi},
  \citenamefont {Grepstad}, \citenamefont {Tybell},\ and\ \citenamefont
  {Matsumoto}}]{Tanaka2010}%
  \BibitemOpen
  \bibfield  {author} {\bibinfo {author} {\bibfnamefont {R.}~\bibnamefont
  {Tanaka}}, \bibinfo {author} {\bibfnamefont {S.}~\bibnamefont {Takata}},
  \bibinfo {author} {\bibfnamefont {M.}~\bibnamefont {Katayama}}, \bibinfo
  {author} {\bibfnamefont {R.}~\bibnamefont {Takahashi}}, \bibinfo {author}
  {\bibfnamefont {J.~K.}\ \bibnamefont {Grepstad}}, \bibinfo {author}
  {\bibfnamefont {T.}~\bibnamefont {Tybell}},\ and\ \bibinfo {author}
  {\bibfnamefont {Y.}~\bibnamefont {Matsumoto}},\ }\href
  {https://doi.org/10.1149/1.3503581} {\bibfield  {journal} {\bibinfo
  {journal} {Journal of The Electrochemical Society}\ }\textbf {\bibinfo
  {volume} {157}},\ \bibinfo {pages} {E181} (\bibinfo {year}
  {2010})}\BibitemShut {NoStop}%
\bibitem [{\citenamefont {Wang}\ \emph {et~al.}(2009)\citenamefont {Wang},
  \citenamefont {Kako},\ and\ \citenamefont {Ye}}]{Wang2009}%
  \BibitemOpen
  \bibfield  {author} {\bibinfo {author} {\bibfnamefont {D.}~\bibnamefont
  {Wang}}, \bibinfo {author} {\bibfnamefont {T.}~\bibnamefont {Kako}},\ and\
  \bibinfo {author} {\bibfnamefont {J.}~\bibnamefont {Ye}},\ }\href
  {https://doi.org/10.1021/jp807393a} {\bibfield  {journal} {\bibinfo
  {journal} {The Journal of Physical Chemistry C}\ }\textbf {\bibinfo {volume}
  {113}},\ \bibinfo {pages} {3785} (\bibinfo {year} {2009})}\BibitemShut
  {NoStop}%
\bibitem [{\citenamefont {Subramanian}\ \emph {et~al.}(2006)\citenamefont
  {Subramanian}, \citenamefont {Roeder},\ and\ \citenamefont
  {Wolf}}]{Subramanian2006}%
  \BibitemOpen
  \bibfield  {author} {\bibinfo {author} {\bibfnamefont {V.}~\bibnamefont
  {Subramanian}}, \bibinfo {author} {\bibfnamefont {R.~K.}\ \bibnamefont
  {Roeder}},\ and\ \bibinfo {author} {\bibfnamefont {E.~E.}\ \bibnamefont
  {Wolf}},\ }\href {https://doi.org/10.1021/ie050693y} {\bibfield  {journal}
  {\bibinfo  {journal} {Industrial \& Engineering Chemistry Research}\ }\textbf
  {\bibinfo {volume} {45}},\ \bibinfo {pages} {2187} (\bibinfo {year}
  {2006})}\BibitemShut {NoStop}%
\bibitem [{\citenamefont {Asai}\ \emph {et~al.}(2014)\citenamefont {Asai},
  \citenamefont {Nemoto}, \citenamefont {Jia}, \citenamefont {Saito},
  \citenamefont {Iwase},\ and\ \citenamefont {Kudo}}]{Asai2014}%
  \BibitemOpen
  \bibfield  {author} {\bibinfo {author} {\bibfnamefont {R.}~\bibnamefont
  {Asai}}, \bibinfo {author} {\bibfnamefont {H.}~\bibnamefont {Nemoto}},
  \bibinfo {author} {\bibfnamefont {Q.}~\bibnamefont {Jia}}, \bibinfo {author}
  {\bibfnamefont {K.}~\bibnamefont {Saito}}, \bibinfo {author} {\bibfnamefont
  {A.}~\bibnamefont {Iwase}},\ and\ \bibinfo {author} {\bibfnamefont
  {A.}~\bibnamefont {Kudo}},\ }\href {https://doi.org/10.1039/C3CC49279F}
  {\bibfield  {journal} {\bibinfo  {journal} {Chem. Commun.}\ }\textbf
  {\bibinfo {volume} {50}},\ \bibinfo {pages} {2543} (\bibinfo {year}
  {2014})}\BibitemShut {NoStop}%
\bibitem [{\citenamefont {Furuhashi}\ \emph {et~al.}(2013)\citenamefont
  {Furuhashi}, \citenamefont {Jia}, \citenamefont {Kudo},\ and\ \citenamefont
  {Onishi}}]{Furuhashi2013}%
  \BibitemOpen
  \bibfield  {author} {\bibinfo {author} {\bibfnamefont {K.}~\bibnamefont
  {Furuhashi}}, \bibinfo {author} {\bibfnamefont {Q.}~\bibnamefont {Jia}},
  \bibinfo {author} {\bibfnamefont {A.}~\bibnamefont {Kudo}},\ and\ \bibinfo
  {author} {\bibfnamefont {H.}~\bibnamefont {Onishi}},\ }\href
  {https://doi.org/10.1021/jp407040p} {\bibfield  {journal} {\bibinfo
  {journal} {The Journal of Physical Chemistry C}\ }\textbf {\bibinfo {volume}
  {117}},\ \bibinfo {pages} {19101} (\bibinfo {year} {2013})}\BibitemShut
  {NoStop}%
\bibitem [{\citenamefont {Jiang}\ \emph {et~al.}(2020)\citenamefont {Jiang},
  \citenamefont {Kato}, \citenamefont {Fujimori}, \citenamefont {Yamakata},\
  and\ \citenamefont {Sakata}}]{Jiang2020}%
  \BibitemOpen
  \bibfield  {author} {\bibinfo {author} {\bibfnamefont {J.}~\bibnamefont
  {Jiang}}, \bibinfo {author} {\bibfnamefont {K.}~\bibnamefont {Kato}},
  \bibinfo {author} {\bibfnamefont {H.}~\bibnamefont {Fujimori}}, \bibinfo
  {author} {\bibfnamefont {A.}~\bibnamefont {Yamakata}},\ and\ \bibinfo
  {author} {\bibfnamefont {Y.}~\bibnamefont {Sakata}},\ }\bibfield  {journal}
  {\bibinfo  {journal} {Journal of Catalysis}\ }\textbf {\bibinfo {volume}
  {390}},\ \href {https://doi.org/https://doi.org/10.1016/j.jcat.2020.07.025}
  {https://doi.org/10.1016/j.jcat.2020.07.025} (\bibinfo {year}
  {2020})\BibitemShut {NoStop}%
\bibitem [{\citenamefont {Zhu}\ \emph {et~al.}(2016)\citenamefont {Zhu},
  \citenamefont {Salvador},\ and\ \citenamefont {Rohrer}}]{Zhu2016}%
  \BibitemOpen
  \bibfield  {author} {\bibinfo {author} {\bibfnamefont {Y.}~\bibnamefont
  {Zhu}}, \bibinfo {author} {\bibfnamefont {P.~A.}\ \bibnamefont {Salvador}},\
  and\ \bibinfo {author} {\bibfnamefont {G.~S.}\ \bibnamefont {Rohrer}},\
  }\href {https://doi.org/10.1021/acs.chemmater.6b02205} {\bibfield  {journal}
  {\bibinfo  {journal} {Chemistry of Materials}\ }\textbf {\bibinfo {volume}
  {28}},\ \bibinfo {pages} {5155} (\bibinfo {year} {2016})}\BibitemShut
  {NoStop}%
\bibitem [{\citenamefont {Zhang}\ \emph {et~al.}(2020)\citenamefont {Zhang},
  \citenamefont {Salvador},\ and\ \citenamefont {Rohrer}}]{Zhang2020}%
  \BibitemOpen
  \bibfield  {author} {\bibinfo {author} {\bibfnamefont {M.}~\bibnamefont
  {Zhang}}, \bibinfo {author} {\bibfnamefont {P.~A.}\ \bibnamefont
  {Salvador}},\ and\ \bibinfo {author} {\bibfnamefont {G.~S.}\ \bibnamefont
  {Rohrer}},\ }\href {https://doi.org/10.1021/acsami.0c04351} {\bibfield
  {journal} {\bibinfo  {journal} {ACS Applied Materials {\&} Interfaces}\
  }\textbf {\bibinfo {volume} {12}},\ \bibinfo {pages} {23617} (\bibinfo {year}
  {2020})}\BibitemShut {NoStop}%
\bibitem [{\citenamefont {Schultz}\ \emph {et~al.}(2011)\citenamefont
  {Schultz}, \citenamefont {Zhang}, \citenamefont {Salvador},\ and\
  \citenamefont {Rohrer}}]{Schultz2011}%
  \BibitemOpen
  \bibfield  {author} {\bibinfo {author} {\bibfnamefont {A.~M.}\ \bibnamefont
  {Schultz}}, \bibinfo {author} {\bibfnamefont {Y.}~\bibnamefont {Zhang}},
  \bibinfo {author} {\bibfnamefont {P.~A.}\ \bibnamefont {Salvador}},\ and\
  \bibinfo {author} {\bibfnamefont {G.~S.}\ \bibnamefont {Rohrer}},\ }\href
  {https://doi.org/10.1021/am200127c} {\bibfield  {journal} {\bibinfo
  {journal} {ACS Applied Materials {\&} Interfaces}\ }\textbf {\bibinfo
  {volume} {3}},\ \bibinfo {pages} {1562} (\bibinfo {year} {2011})}\BibitemShut
  {NoStop}%
\bibitem [{\citenamefont {Ham}\ \emph {et~al.}(2016)\citenamefont {Ham},
  \citenamefont {Hisatomi}, \citenamefont {Goto}, \citenamefont {Moriya},
  \citenamefont {Sakata}, \citenamefont {Yamakata}, \citenamefont {Kubota},\
  and\ \citenamefont {Domen}}]{Ham2016}%
  \BibitemOpen
  \bibfield  {author} {\bibinfo {author} {\bibfnamefont {Y.}~\bibnamefont
  {Ham}}, \bibinfo {author} {\bibfnamefont {T.}~\bibnamefont {Hisatomi}},
  \bibinfo {author} {\bibfnamefont {Y.}~\bibnamefont {Goto}}, \bibinfo {author}
  {\bibfnamefont {Y.}~\bibnamefont {Moriya}}, \bibinfo {author} {\bibfnamefont
  {Y.}~\bibnamefont {Sakata}}, \bibinfo {author} {\bibfnamefont
  {A.}~\bibnamefont {Yamakata}}, \bibinfo {author} {\bibfnamefont
  {J.}~\bibnamefont {Kubota}},\ and\ \bibinfo {author} {\bibfnamefont
  {K.}~\bibnamefont {Domen}},\ }\href@noop {} {\bibfield  {journal} {\bibinfo
  {journal} {Journal of Materials Chemistry A}\ }\textbf {\bibinfo {volume}
  {4}},\ \bibinfo {pages} {3027} (\bibinfo {year} {2016})}\BibitemShut
  {NoStop}%
\bibitem [{\citenamefont {Shen}\ \emph {et~al.}(2010)\citenamefont {Shen},
  \citenamefont {Small}, \citenamefont {Wang}, \citenamefont {Allen},
  \citenamefont {Fern\'{a}ndez-Serra}, \citenamefont {Hybertsen},\ and\
  \citenamefont {Muckerman}}]{Shen2010}%
  \BibitemOpen
  \bibfield  {author} {\bibinfo {author} {\bibfnamefont {X.}~\bibnamefont
  {Shen}}, \bibinfo {author} {\bibfnamefont {Y.~A.}\ \bibnamefont {Small}},
  \bibinfo {author} {\bibfnamefont {J.}~\bibnamefont {Wang}}, \bibinfo {author}
  {\bibfnamefont {P.~B.}\ \bibnamefont {Allen}}, \bibinfo {author}
  {\bibfnamefont {M.~V.}\ \bibnamefont {Fern\'{a}ndez-Serra}}, \bibinfo
  {author} {\bibfnamefont {M.~S.}\ \bibnamefont {Hybertsen}},\ and\ \bibinfo
  {author} {\bibfnamefont {J.~T.}\ \bibnamefont {Muckerman}},\ }\href
  {https://doi.org/10.1021/jp102958s} {\bibfield  {journal} {\bibinfo
  {journal} {The Journal of Physical Chemistry C}\ }\textbf {\bibinfo {volume}
  {114}},\ \bibinfo {pages} {13695} (\bibinfo {year} {2010})}\BibitemShut
  {NoStop}%
\bibitem [{\citenamefont {Selcuk}\ and\ \citenamefont
  {Selloni}(2016)}]{Selcuk2016}%
  \BibitemOpen
  \bibfield  {author} {\bibinfo {author} {\bibfnamefont {S.}~\bibnamefont
  {Selcuk}}\ and\ \bibinfo {author} {\bibfnamefont {A.}~\bibnamefont
  {Selloni}},\ }\href@noop {} {\bibfield  {journal} {\bibinfo  {journal}
  {Nature Materials}\ }\textbf {\bibinfo {volume} {15}},\ \bibinfo {pages}
  {1107} (\bibinfo {year} {2016})}\BibitemShut {NoStop}%
\bibitem [{\citenamefont {Crespillo}\ \emph {et~al.}(2019)\citenamefont
  {Crespillo}, \citenamefont {Graham}, \citenamefont {Agull\'{o}-L\'{o}pez},
  \citenamefont {Zhang},\ and\ \citenamefont {Weber}}]{Crespillo2019}%
  \BibitemOpen
  \bibfield  {author} {\bibinfo {author} {\bibfnamefont {M.}~\bibnamefont
  {Crespillo}}, \bibinfo {author} {\bibfnamefont {J.}~\bibnamefont {Graham}},
  \bibinfo {author} {\bibfnamefont {F.}~\bibnamefont {Agull\'{o}-L\'{o}pez}},
  \bibinfo {author} {\bibfnamefont {Y.}~\bibnamefont {Zhang}},\ and\ \bibinfo
  {author} {\bibfnamefont {W.}~\bibnamefont {Weber}},\ }\href
  {https://doi.org/10.3390/cryst9020095} {\bibfield  {journal} {\bibinfo
  {journal} {Crystals}\ }\textbf {\bibinfo {volume} {9}},\ \bibinfo {pages}
  {95} (\bibinfo {year} {2019})}\BibitemShut {NoStop}%
\bibitem [{\citenamefont {Kawasaki}\ \emph {et~al.}(1994)\citenamefont
  {Kawasaki}, \citenamefont {Takahashi}, \citenamefont {Maeda}, \citenamefont
  {Tsuchiya}, \citenamefont {Shinohara}, \citenamefont {Ishiyama},
  \citenamefont {Yonezawa}, \citenamefont {Yoshimoto},\ and\ \citenamefont
  {Koinuma}}]{kawasaki1994atomic}%
  \BibitemOpen
  \bibfield  {author} {\bibinfo {author} {\bibfnamefont {M.}~\bibnamefont
  {Kawasaki}}, \bibinfo {author} {\bibfnamefont {K.}~\bibnamefont {Takahashi}},
  \bibinfo {author} {\bibfnamefont {T.}~\bibnamefont {Maeda}}, \bibinfo
  {author} {\bibfnamefont {R.}~\bibnamefont {Tsuchiya}}, \bibinfo {author}
  {\bibfnamefont {M.}~\bibnamefont {Shinohara}}, \bibinfo {author}
  {\bibfnamefont {O.}~\bibnamefont {Ishiyama}}, \bibinfo {author}
  {\bibfnamefont {T.}~\bibnamefont {Yonezawa}}, \bibinfo {author}
  {\bibfnamefont {M.}~\bibnamefont {Yoshimoto}},\ and\ \bibinfo {author}
  {\bibfnamefont {H.}~\bibnamefont {Koinuma}},\ }\href@noop {} {\bibfield
  {journal} {\bibinfo  {journal} {Science}\ }\textbf {\bibinfo {volume}
  {266}},\ \bibinfo {pages} {1540} (\bibinfo {year} {1994})}\BibitemShut
  {NoStop}%
\bibitem [{\citenamefont {Koster}\ \emph {et~al.}(1998)\citenamefont {Koster},
  \citenamefont {Kropman}, \citenamefont {Rijnders}, \citenamefont {Blank},\
  and\ \citenamefont {Rogalla}}]{koster1998quasi}%
  \BibitemOpen
  \bibfield  {author} {\bibinfo {author} {\bibfnamefont {G.}~\bibnamefont
  {Koster}}, \bibinfo {author} {\bibfnamefont {B.~L.}\ \bibnamefont {Kropman}},
  \bibinfo {author} {\bibfnamefont {G.~J.}\ \bibnamefont {Rijnders}}, \bibinfo
  {author} {\bibfnamefont {D.~H.}\ \bibnamefont {Blank}},\ and\ \bibinfo
  {author} {\bibfnamefont {H.}~\bibnamefont {Rogalla}},\ }\href@noop {}
  {\bibfield  {journal} {\bibinfo  {journal} {Applied Physics Letters}\
  }\textbf {\bibinfo {volume} {73}},\ \bibinfo {pages} {2920} (\bibinfo {year}
  {1998})}\BibitemShut {NoStop}%
\bibitem [{\citenamefont {Ohnishi}\ \emph {et~al.}(2004)\citenamefont
  {Ohnishi}, \citenamefont {Shibuya}, \citenamefont {Lippmaa}, \citenamefont
  {Kobayashi}, \citenamefont {Kumigashira}, \citenamefont {Oshima},\ and\
  \citenamefont {Koinuma}}]{ohnishi2004preparation}%
  \BibitemOpen
  \bibfield  {author} {\bibinfo {author} {\bibfnamefont {T.}~\bibnamefont
  {Ohnishi}}, \bibinfo {author} {\bibfnamefont {K.}~\bibnamefont {Shibuya}},
  \bibinfo {author} {\bibfnamefont {M.}~\bibnamefont {Lippmaa}}, \bibinfo
  {author} {\bibfnamefont {D.}~\bibnamefont {Kobayashi}}, \bibinfo {author}
  {\bibfnamefont {H.}~\bibnamefont {Kumigashira}}, \bibinfo {author}
  {\bibfnamefont {M.}~\bibnamefont {Oshima}},\ and\ \bibinfo {author}
  {\bibfnamefont {H.}~\bibnamefont {Koinuma}},\ }\href@noop {} {\bibfield
  {journal} {\bibinfo  {journal} {Applied physics letters}\ }\textbf {\bibinfo
  {volume} {85}},\ \bibinfo {pages} {272} (\bibinfo {year} {2004})}\BibitemShut
  {NoStop}%
\bibitem [{\citenamefont {Velasco-Davalos}\ \emph {et~al.}(2013)\citenamefont
  {Velasco-Davalos}, \citenamefont {Thomas},\ and\ \citenamefont
  {Ruediger}}]{Velasco2013}%
  \BibitemOpen
  \bibfield  {author} {\bibinfo {author} {\bibfnamefont {I.}~\bibnamefont
  {Velasco-Davalos}}, \bibinfo {author} {\bibfnamefont {R.}~\bibnamefont
  {Thomas}},\ and\ \bibinfo {author} {\bibfnamefont {A.}~\bibnamefont
  {Ruediger}},\ }\href {https://doi.org/10.1063/1.4831681} {\bibfield
  {journal} {\bibinfo  {journal} {Applied Physics Letters}\ }\textbf {\bibinfo
  {volume} {103}},\ \bibinfo {pages} {202905} (\bibinfo {year}
  {2013})}\BibitemShut {NoStop}%
\bibitem [{\citenamefont {Chambers}\ \emph {et~al.}(2012)\citenamefont
  {Chambers}, \citenamefont {Droubay}, \citenamefont {Capan},\ and\
  \citenamefont {Sun}}]{Chambers2012}%
  \BibitemOpen
  \bibfield  {author} {\bibinfo {author} {\bibfnamefont {S.}~\bibnamefont
  {Chambers}}, \bibinfo {author} {\bibfnamefont {T.}~\bibnamefont {Droubay}},
  \bibinfo {author} {\bibfnamefont {C.}~\bibnamefont {Capan}},\ and\ \bibinfo
  {author} {\bibfnamefont {G.}~\bibnamefont {Sun}},\ }\href
  {https://doi.org/https://doi.org/10.1016/j.susc.2011.11.029} {\bibfield
  {journal} {\bibinfo  {journal} {Surface Science}\ }\textbf {\bibinfo {volume}
  {606}},\ \bibinfo {pages} {554 } (\bibinfo {year} {2012})}\BibitemShut
  {NoStop}%
\bibitem [{\citenamefont {Bachelet}\ \emph
  {et~al.}(2009{\natexlab{a}})\citenamefont {Bachelet}, \citenamefont
  {S\'anchez}, \citenamefont {Palomares}, \citenamefont {Ocal},\ and\
  \citenamefont {Fontcuberta}}]{Bachelet2009}%
  \BibitemOpen
  \bibfield  {author} {\bibinfo {author} {\bibfnamefont {R.}~\bibnamefont
  {Bachelet}}, \bibinfo {author} {\bibfnamefont {F.}~\bibnamefont {S\'anchez}},
  \bibinfo {author} {\bibfnamefont {F.~J.}\ \bibnamefont {Palomares}}, \bibinfo
  {author} {\bibfnamefont {C.}~\bibnamefont {Ocal}},\ and\ \bibinfo {author}
  {\bibfnamefont {J.}~\bibnamefont {Fontcuberta}},\ }\href
  {https://doi.org/10.1063/1.3240869} {\bibfield  {journal} {\bibinfo
  {journal} {Applied Physics Letters}\ }\textbf {\bibinfo {volume} {95}},\
  \bibinfo {pages} {141915} (\bibinfo {year} {2009}{\natexlab{a}})}\BibitemShut
  {NoStop}%
\bibitem [{\citenamefont {Tiwari}\ and\ \citenamefont
  {Dunn}(2009)}]{Tiwari2009}%
  \BibitemOpen
  \bibfield  {author} {\bibinfo {author} {\bibfnamefont {D.}~\bibnamefont
  {Tiwari}}\ and\ \bibinfo {author} {\bibfnamefont {S.}~\bibnamefont {Dunn}},\
  }\href {https://doi.org/10.1007/s10853-009-3472-1} {\bibfield  {journal}
  {\bibinfo  {journal} {Journal of Materials Science}\ }\textbf {\bibinfo
  {volume} {44}},\ \bibinfo {pages} {5063} (\bibinfo {year}
  {2009})}\BibitemShut {NoStop}%
\bibitem [{\citenamefont {Szot}\ and\ \citenamefont
  {Speier}(1999)}]{szot1999surfaces}%
  \BibitemOpen
  \bibfield  {author} {\bibinfo {author} {\bibfnamefont {K.}~\bibnamefont
  {Szot}}\ and\ \bibinfo {author} {\bibfnamefont {W.}~\bibnamefont {Speier}},\
  }\href@noop {} {\bibfield  {journal} {\bibinfo  {journal} {Physical Review
  B}\ }\textbf {\bibinfo {volume} {60}},\ \bibinfo {pages} {5909} (\bibinfo
  {year} {1999})}\BibitemShut {NoStop}%
\bibitem [{\citenamefont {Bachelet}\ \emph
  {et~al.}(2009{\natexlab{b}})\citenamefont {Bachelet}, \citenamefont
  {S{\'a}nchez}, \citenamefont {Santiso}, \citenamefont {Munuera},
  \citenamefont {Ocal},\ and\ \citenamefont {Fontcuberta}}]{bachelet2009self}%
  \BibitemOpen
  \bibfield  {author} {\bibinfo {author} {\bibfnamefont {R.}~\bibnamefont
  {Bachelet}}, \bibinfo {author} {\bibfnamefont {F.}~\bibnamefont
  {S{\'a}nchez}}, \bibinfo {author} {\bibfnamefont {J.}~\bibnamefont
  {Santiso}}, \bibinfo {author} {\bibfnamefont {C.}~\bibnamefont {Munuera}},
  \bibinfo {author} {\bibfnamefont {C.}~\bibnamefont {Ocal}},\ and\ \bibinfo
  {author} {\bibfnamefont {J.}~\bibnamefont {Fontcuberta}},\ }\href@noop {}
  {\bibfield  {journal} {\bibinfo  {journal} {Chemistry of materials}\ }\textbf
  {\bibinfo {volume} {21}},\ \bibinfo {pages} {2494} (\bibinfo {year}
  {2009}{\natexlab{b}})}\BibitemShut {NoStop}%
\bibitem [{\citenamefont {Domingo}\ \emph {et~al.}(2019)\citenamefont
  {Domingo}, \citenamefont {Pach}, \citenamefont {Cordero-Edwards},
  \citenamefont {P{\'e}rez-Dieste}, \citenamefont {Escudero},\ and\
  \citenamefont {Verdaguer}}]{domingo2019water}%
  \BibitemOpen
  \bibfield  {author} {\bibinfo {author} {\bibfnamefont {N.}~\bibnamefont
  {Domingo}}, \bibinfo {author} {\bibfnamefont {E.}~\bibnamefont {Pach}},
  \bibinfo {author} {\bibfnamefont {K.}~\bibnamefont {Cordero-Edwards}},
  \bibinfo {author} {\bibfnamefont {V.}~\bibnamefont {P{\'e}rez-Dieste}},
  \bibinfo {author} {\bibfnamefont {C.}~\bibnamefont {Escudero}},\ and\
  \bibinfo {author} {\bibfnamefont {A.}~\bibnamefont {Verdaguer}},\ }\href@noop
  {} {\bibfield  {journal} {\bibinfo  {journal} {Physical Chemistry Chemical
  Physics}\ }\textbf {\bibinfo {volume} {21}},\ \bibinfo {pages} {4920}
  (\bibinfo {year} {2019})}\BibitemShut {NoStop}%
\bibitem [{\citenamefont {Wang}\ \emph {et~al.}(2012)\citenamefont {Wang},
  \citenamefont {Pedroza}, \citenamefont {Poissier},\ and\ \citenamefont
  {Fern\'{a}ndez-Serra}}]{Wang2012}%
  \BibitemOpen
  \bibfield  {author} {\bibinfo {author} {\bibfnamefont {J.}~\bibnamefont
  {Wang}}, \bibinfo {author} {\bibfnamefont {L.~S.}\ \bibnamefont {Pedroza}},
  \bibinfo {author} {\bibfnamefont {A.}~\bibnamefont {Poissier}},\ and\
  \bibinfo {author} {\bibfnamefont {M.~V.}\ \bibnamefont
  {Fern\'{a}ndez-Serra}},\ }\href {https://doi.org/10.1021/jp302793s}
  {\bibfield  {journal} {\bibinfo  {journal} {The Journal of Physical Chemistry
  C}\ }\textbf {\bibinfo {volume} {116}},\ \bibinfo {pages} {14382} (\bibinfo
  {year} {2012})}\BibitemShut {NoStop}%
\bibitem [{\citenamefont {Kharche}\ \emph {et~al.}(2014)\citenamefont
  {Kharche}, \citenamefont {Muckerman},\ and\ \citenamefont
  {Hybertsen}}]{Kharche2014}%
  \BibitemOpen
  \bibfield  {author} {\bibinfo {author} {\bibfnamefont {N.}~\bibnamefont
  {Kharche}}, \bibinfo {author} {\bibfnamefont {J.~T.}\ \bibnamefont
  {Muckerman}},\ and\ \bibinfo {author} {\bibfnamefont {M.~S.}\ \bibnamefont
  {Hybertsen}},\ }\href@noop {} {\bibfield  {journal} {\bibinfo  {journal}
  {Physical review letters}\ }\textbf {\bibinfo {volume} {113}},\ \bibinfo
  {pages} {176802} (\bibinfo {year} {2014})}\BibitemShut {NoStop}%
\bibitem [{\citenamefont {Zhong}\ and\ \citenamefont
  {Hansmann}(2016)}]{Zhong2016}%
  \BibitemOpen
  \bibfield  {author} {\bibinfo {author} {\bibfnamefont {Z.}~\bibnamefont
  {Zhong}}\ and\ \bibinfo {author} {\bibfnamefont {P.}~\bibnamefont
  {Hansmann}},\ }\href {https://doi.org/10.1103/PhysRevB.93.235116} {\bibfield
  {journal} {\bibinfo  {journal} {Phys. Rev. B}\ }\textbf {\bibinfo {volume}
  {93}},\ \bibinfo {pages} {235116} (\bibinfo {year} {2016})}\BibitemShut
  {NoStop}%
\bibitem [{\citenamefont {Sung}\ \emph {et~al.}(2020)\citenamefont {Sung},
  \citenamefont {Mochizuki},\ and\ \citenamefont {Oba}}]{Sung2020}%
  \BibitemOpen
  \bibfield  {author} {\bibinfo {author} {\bibfnamefont {H.-J.}\ \bibnamefont
  {Sung}}, \bibinfo {author} {\bibfnamefont {Y.}~\bibnamefont {Mochizuki}},\
  and\ \bibinfo {author} {\bibfnamefont {F.}~\bibnamefont {Oba}},\ }\href
  {https://doi.org/10.1103/PhysRevMaterials.4.044606} {\bibfield  {journal}
  {\bibinfo  {journal} {Phys. Rev. Materials}\ }\textbf {\bibinfo {volume}
  {4}},\ \bibinfo {pages} {044606} (\bibinfo {year} {2020})}\BibitemShut
  {NoStop}%
\bibitem [{\citenamefont {Ma}\ \emph {et~al.}(2020)\citenamefont {Ma},
  \citenamefont {Jacobs}, \citenamefont {Booske},\ and\ \citenamefont
  {Morgan}}]{Ma2020}%
  \BibitemOpen
  \bibfield  {author} {\bibinfo {author} {\bibfnamefont {T.}~\bibnamefont
  {Ma}}, \bibinfo {author} {\bibfnamefont {R.}~\bibnamefont {Jacobs}}, \bibinfo
  {author} {\bibfnamefont {J.}~\bibnamefont {Booske}},\ and\ \bibinfo {author}
  {\bibfnamefont {D.}~\bibnamefont {Morgan}},\ }\href
  {https://doi.org/10.1063/1.5143325} {\bibfield  {journal} {\bibinfo
  {journal} {APL Materials}\ }\textbf {\bibinfo {volume} {8}},\ \bibinfo
  {pages} {071110} (\bibinfo {year} {2020})}\BibitemShut {NoStop}%
\bibitem [{\citenamefont {Ricca}\ \emph {et~al.}(2020)\citenamefont {Ricca},
  \citenamefont {Timrov}, \citenamefont {Cococcioni}, \citenamefont {Marzari},\
  and\ \citenamefont {Aschauer}}]{Marzari2020}%
  \BibitemOpen
  \bibfield  {author} {\bibinfo {author} {\bibfnamefont {C.}~\bibnamefont
  {Ricca}}, \bibinfo {author} {\bibfnamefont {I.}~\bibnamefont {Timrov}},
  \bibinfo {author} {\bibfnamefont {M.}~\bibnamefont {Cococcioni}}, \bibinfo
  {author} {\bibfnamefont {N.}~\bibnamefont {Marzari}},\ and\ \bibinfo {author}
  {\bibfnamefont {U.}~\bibnamefont {Aschauer}},\ }\href
  {https://doi.org/10.1103/PhysRevResearch.2.023313} {\bibfield  {journal}
  {\bibinfo  {journal} {Phys. Rev. Research}\ }\textbf {\bibinfo {volume}
  {2}},\ \bibinfo {pages} {023313} (\bibinfo {year} {2020})}\BibitemShut
  {NoStop}%
\bibitem [{\citenamefont {Kim}\ \emph {et~al.}(2015)\citenamefont {Kim},
  \citenamefont {Ping}, \citenamefont {Galli},\ and\ \citenamefont
  {Choi}}]{Kim2015}%
  \BibitemOpen
  \bibfield  {author} {\bibinfo {author} {\bibfnamefont {T.~W.}\ \bibnamefont
  {Kim}}, \bibinfo {author} {\bibfnamefont {Y.}~\bibnamefont {Ping}}, \bibinfo
  {author} {\bibfnamefont {G.~A.}\ \bibnamefont {Galli}},\ and\ \bibinfo
  {author} {\bibfnamefont {K.-S.}\ \bibnamefont {Choi}},\ }\href
  {https://doi.org/10.1038/ncomms9769} {\bibfield  {journal} {\bibinfo
  {journal} {Nature Communications}\ }\textbf {\bibinfo {volume} {6}},\
  \bibinfo {pages} {8769} (\bibinfo {year} {2015})}\BibitemShut {NoStop}%
\bibitem [{\citenamefont {Heyd}\ \emph {et~al.}(2003)\citenamefont {Heyd},
  \citenamefont {Scuseria},\ and\ \citenamefont {Ernzerhof}}]{HSE1}%
  \BibitemOpen
  \bibfield  {author} {\bibinfo {author} {\bibfnamefont {J.}~\bibnamefont
  {Heyd}}, \bibinfo {author} {\bibfnamefont {G.~E.}\ \bibnamefont {Scuseria}},\
  and\ \bibinfo {author} {\bibfnamefont {M.}~\bibnamefont {Ernzerhof}},\ }\href
  {https://doi.org/10.1063/1.1564060} {\bibfield  {journal} {\bibinfo
  {journal} {The Journal of Chemical Physics}\ }\textbf {\bibinfo {volume}
  {118}},\ \bibinfo {pages} {8207} (\bibinfo {year} {2003})}\BibitemShut
  {NoStop}%
\bibitem [{\citenamefont {Heyd}\ \emph {et~al.}(2006)\citenamefont {Heyd},
  \citenamefont {Scuseria},\ and\ \citenamefont {Ernzerhof}}]{HSE2}%
  \BibitemOpen
  \bibfield  {author} {\bibinfo {author} {\bibfnamefont {J.}~\bibnamefont
  {Heyd}}, \bibinfo {author} {\bibfnamefont {G.~E.}\ \bibnamefont {Scuseria}},\
  and\ \bibinfo {author} {\bibfnamefont {M.}~\bibnamefont {Ernzerhof}},\ }\href
  {https://doi.org/10.1063/1.2204597} {\bibfield  {journal} {\bibinfo
  {journal} {The Journal of Chemical Physics}\ }\textbf {\bibinfo {volume}
  {124}},\ \bibinfo {pages} {219906} (\bibinfo {year} {2006})}\BibitemShut
  {NoStop}%
\bibitem [{\citenamefont {Li}\ \emph {et~al.}(2011)\citenamefont {Li},
  \citenamefont {Muckerman}, \citenamefont {Hybertsen},\ and\ \citenamefont
  {Allen}}]{PhysRevB.83.134202}%
  \BibitemOpen
  \bibfield  {author} {\bibinfo {author} {\bibfnamefont {L.}~\bibnamefont
  {Li}}, \bibinfo {author} {\bibfnamefont {J.~T.}\ \bibnamefont {Muckerman}},
  \bibinfo {author} {\bibfnamefont {M.~S.}\ \bibnamefont {Hybertsen}},\ and\
  \bibinfo {author} {\bibfnamefont {P.~B.}\ \bibnamefont {Allen}},\ }\href
  {https://doi.org/10.1103/PhysRevB.83.134202} {\bibfield  {journal} {\bibinfo
  {journal} {Phys. Rev. B}\ }\textbf {\bibinfo {volume} {83}},\ \bibinfo
  {pages} {134202} (\bibinfo {year} {2011})}\BibitemShut {NoStop}%
\bibitem [{\citenamefont {Hammes-Schiffer}\ and\ \citenamefont
  {Galli}(2021)}]{Hammes-Schiffer2021}%
  \BibitemOpen
  \bibfield  {author} {\bibinfo {author} {\bibfnamefont {S.}~\bibnamefont
  {Hammes-Schiffer}}\ and\ \bibinfo {author} {\bibfnamefont {G.}~\bibnamefont
  {Galli}},\ }\href {https://doi.org/10.1038/s41560-021-00827-4} {\bibfield
  {journal} {\bibinfo  {journal} {Nature Energy}\ }\textbf {\bibinfo {volume}
  {6}},\ \bibinfo {pages} {700} (\bibinfo {year} {2021})}\BibitemShut {NoStop}%
\bibitem [{\citenamefont {Reyes-Lillo}\ \emph {et~al.}(2016)\citenamefont
  {Reyes-Lillo}, \citenamefont {Rangel}, \citenamefont {Bruneval},\ and\
  \citenamefont {Neaton}}]{PhysRevB.94.041107}%
  \BibitemOpen
  \bibfield  {author} {\bibinfo {author} {\bibfnamefont {S.~E.}\ \bibnamefont
  {Reyes-Lillo}}, \bibinfo {author} {\bibfnamefont {T.}~\bibnamefont {Rangel}},
  \bibinfo {author} {\bibfnamefont {F.}~\bibnamefont {Bruneval}},\ and\
  \bibinfo {author} {\bibfnamefont {J.~B.}\ \bibnamefont {Neaton}},\ }\href
  {https://doi.org/10.1103/PhysRevB.94.041107} {\bibfield  {journal} {\bibinfo
  {journal} {Phys. Rev. B}\ }\textbf {\bibinfo {volume} {94}},\ \bibinfo
  {pages} {041107} (\bibinfo {year} {2016})}\BibitemShut {NoStop}%
\bibitem [{\citenamefont {Car}(2016)}]{Car2016}%
  \BibitemOpen
  \bibfield  {author} {\bibinfo {author} {\bibfnamefont {R.}~\bibnamefont
  {Car}},\ }\href {https://doi.org/10.1038/nchem.2605} {\bibfield  {journal}
  {\bibinfo  {journal} {Nature Chemistry}\ }\textbf {\bibinfo {volume} {8}},\
  \bibinfo {pages} {820} (\bibinfo {year} {2016})}\BibitemShut {NoStop}%
\bibitem [{\citenamefont {Kresse}\ and\ \citenamefont
  {Furthm\"uller}(1996)}]{Kresse1996}%
  \BibitemOpen
  \bibfield  {author} {\bibinfo {author} {\bibfnamefont {G.}~\bibnamefont
  {Kresse}}\ and\ \bibinfo {author} {\bibfnamefont {J.}~\bibnamefont
  {Furthm\"uller}},\ }\href {https://doi.org/10.1103/PhysRevB.54.11169}
  {\bibfield  {journal} {\bibinfo  {journal} {Phys. Rev. B}\ }\textbf {\bibinfo
  {volume} {54}},\ \bibinfo {pages} {11169} (\bibinfo {year}
  {1996})}\BibitemShut {NoStop}%
\bibitem [{\citenamefont {Bl\"ochl}(1994)}]{PAW}%
  \BibitemOpen
  \bibfield  {author} {\bibinfo {author} {\bibfnamefont {P.~E.}\ \bibnamefont
  {Bl\"ochl}},\ }\href {https://doi.org/10.1103/PhysRevB.50.17953} {\bibfield
  {journal} {\bibinfo  {journal} {Phys. Rev. B}\ }\textbf {\bibinfo {volume}
  {50}},\ \bibinfo {pages} {17953} (\bibinfo {year} {1994})}\BibitemShut
  {NoStop}%
\bibitem [{\citenamefont {Berland}\ and\ \citenamefont
  {Hyldgaard}(2014)}]{BH1989}%
  \BibitemOpen
  \bibfield  {author} {\bibinfo {author} {\bibfnamefont {K.}~\bibnamefont
  {Berland}}\ and\ \bibinfo {author} {\bibfnamefont {P.}~\bibnamefont
  {Hyldgaard}},\ }\href {https://doi.org/10.1103/PhysRevB.89.035412} {\bibfield
   {journal} {\bibinfo  {journal} {Phys. Rev. B}\ }\textbf {\bibinfo {volume}
  {89}},\ \bibinfo {pages} {035412} (\bibinfo {year} {2014})}\BibitemShut
  {NoStop}%
\bibitem [{\citenamefont {Sharma}\ and\ \citenamefont
  {Fern\'andez-Serra}(2020)}]{Sharma2020}%
  \BibitemOpen
  \bibfield  {author} {\bibinfo {author} {\bibfnamefont {V.}~\bibnamefont
  {Sharma}}\ and\ \bibinfo {author} {\bibfnamefont {M.}~\bibnamefont
  {Fern\'andez-Serra}},\ }\href
  {https://doi.org/10.1103/PhysRevResearch.2.043082} {\bibfield  {journal}
  {\bibinfo  {journal} {Phys. Rev. Research}\ }\textbf {\bibinfo {volume}
  {2}},\ \bibinfo {pages} {043082} (\bibinfo {year} {2020})}\BibitemShut
  {NoStop}%
\bibitem [{\citenamefont {Takanabe}(2017)}]{Takanabe2017}%
  \BibitemOpen
  \bibfield  {author} {\bibinfo {author} {\bibfnamefont {K.}~\bibnamefont
  {Takanabe}},\ }\href {https://doi.org/10.1021/acscatal.7b02662} {\bibfield
  {journal} {\bibinfo  {journal} {ACS Catalysis}\ }\textbf {\bibinfo {volume}
  {7}},\ \bibinfo {pages} {8006} (\bibinfo {year} {2017})}\BibitemShut
  {NoStop}%
\bibitem [{\citenamefont {Li}\ and\ \citenamefont {Wu}(2015)}]{Li2015}%
  \BibitemOpen
  \bibfield  {author} {\bibinfo {author} {\bibfnamefont {J.}~\bibnamefont
  {Li}}\ and\ \bibinfo {author} {\bibfnamefont {N.}~\bibnamefont {Wu}},\ }\href
  {https://doi.org/10.1039/C4CY00974F} {\bibfield  {journal} {\bibinfo
  {journal} {Catal. Sci. Technol.}\ }\textbf {\bibinfo {volume} {5}},\ \bibinfo
  {pages} {1360} (\bibinfo {year} {2015})}\BibitemShut {NoStop}%
\bibitem [{\citenamefont {III}(2020)}]{NIST}%
  \BibitemOpen
  \bibfield  {author} {\bibinfo {author} {\bibfnamefont {R.~D.~J.}\
  \bibnamefont {III}},\ }\href {https://doi.org/10.18434/T47C7Z} {\bibinfo
  {title} {{NIST Computational Chemistry Comparison and Benchmark Database}}}
  (\bibinfo {year} {2020})\BibitemShut {NoStop}%
\bibitem [{\citenamefont {Bartmess}(1994)}]{Bartmess1994}%
  \BibitemOpen
  \bibfield  {author} {\bibinfo {author} {\bibfnamefont {J.~E.}\ \bibnamefont
  {Bartmess}},\ }\href {https://doi.org/10.1021/j100076a029} {\bibfield
  {journal} {\bibinfo  {journal} {The Journal of Physical Chemistry}\ }\textbf
  {\bibinfo {volume} {98}},\ \bibinfo {pages} {6420} (\bibinfo {year}
  {1994})}\BibitemShut {NoStop}%
\bibitem [{\citenamefont {McQuarrie}(1975)}]{McQuarrie1975}%
  \BibitemOpen
  \bibfield  {author} {\bibinfo {author} {\bibfnamefont {D.~A. D.~A.}\
  \bibnamefont {McQuarrie}},\ }\href@noop {} {\emph {\bibinfo {title}
  {Statistical mechanics / Donald A. McQuarrie}}},\ Harper's chemistry series\
  (\bibinfo  {publisher} {Harper {\&} Row},\ \bibinfo {address} {New York},\
  \bibinfo {year} {1975})\ \bibinfo {note} {``Portions of this work were
  originally published under the title: Statistical
  thermodynamics.''}\BibitemShut {NoStop}%
\bibitem [{\citenamefont {Tissandier}\ \emph {et~al.}(1998)\citenamefont
  {Tissandier}, \citenamefont {Cowen}, \citenamefont {Feng}, \citenamefont
  {Gundlach}, \citenamefont {Cohen}, \citenamefont {Earhart}, \citenamefont
  {Coe},\ and\ \citenamefont {Tuttle}}]{Tissandier1998}%
  \BibitemOpen
  \bibfield  {author} {\bibinfo {author} {\bibfnamefont {M.~D.}\ \bibnamefont
  {Tissandier}}, \bibinfo {author} {\bibfnamefont {K.~A.}\ \bibnamefont
  {Cowen}}, \bibinfo {author} {\bibfnamefont {W.~Y.}\ \bibnamefont {Feng}},
  \bibinfo {author} {\bibfnamefont {E.}~\bibnamefont {Gundlach}}, \bibinfo
  {author} {\bibfnamefont {M.~H.}\ \bibnamefont {Cohen}}, \bibinfo {author}
  {\bibfnamefont {A.~D.}\ \bibnamefont {Earhart}}, \bibinfo {author}
  {\bibfnamefont {J.~V.}\ \bibnamefont {Coe}},\ and\ \bibinfo {author}
  {\bibfnamefont {T.~R.}\ \bibnamefont {Tuttle}},\ }\href
  {https://doi.org/10.1021/jp982638r} {\bibfield  {journal} {\bibinfo
  {journal} {The Journal of Physical Chemistry A}\ }\textbf {\bibinfo {volume}
  {102}},\ \bibinfo {pages} {7787} (\bibinfo {year} {1998})}\BibitemShut
  {NoStop}%
\bibitem [{\citenamefont {Sharma}(2021)}]{vidushithesis}%
  \BibitemOpen
  \bibfield  {author} {\bibinfo {author} {\bibfnamefont {V.}~\bibnamefont
  {Sharma}},\ }\emph {\bibinfo {title} {Density Functional Theory Studies of
  Water Structure and Photo-Reactivity in Oxide Materials}},\ \href
  {https://www.proquest.com/openview/852f3768f49905a5d3c0c5e85036220a}
  {\bibinfo {type} {{Ph.D. Thesis}}} (\bibinfo {year} {2021})\BibitemShut
  {NoStop}%
\bibitem [{dat()}]{data}%
  \BibitemOpen
  \href {https://doi.org/10.5281/zenodo.5838976} {\bibinfo {title} {Zenodo data
  repository}}\BibitemShut {NoStop}%
\end{thebibliography}%

%%%%%%%%%%%%%%%%%%%%%%%%%%%%%%%%%%%%%%%%%%%%%%%%%%%%%%%%%%%%%%%%%%%
%%%              SUPPLEMENTAL  MATERIAL STARTS HERE             %%%
%%%%%%%%%%%%%%%%%%%%%%%%%%%%%%%%%%%%%%%%%%%%%%%%%%%%%%%%%%%%%%%%%%%

\pagebreak

%reset all style and numbering
\clearpage
\clearpage %needed for two-page reference section
\setcounter{page}{1}
\renewcommand{\thetable}{S\arabic{table}}  
\setcounter{table}{0}
\renewcommand{\thefigure}{S\arabic{figure}}
\setcounter{figure}{0}
\renewcommand{\thesection}{S\arabic{section}}
\setcounter{section}{0}
\renewcommand{\theequation}{S\arabic{equation}}
\setcounter{equation}{0}
\onecolumngrid

%create title
\begin{center}
\textbf{Supplementary Information for\\\vspace{0.5 cm}
\large Photocatalytic water oxidation in SrTiO$_3$ [001] surfaces\\\vspace{0.3 cm}}

Vidushi Sharma$^{1,2,3,4}$, Benjamin Bein$^{1}$, Amanda Lai$^{1}$, Bet\"{u}l Pamuk$^{5}$, Cyrus E. Dreyer$^{1,6}$, Marivi Fern\'{a}ndez-Serra$^{1,2}$ and Matthew Dawber$^{1}$

\small

$^1$\textit{Department of Physics and Astronomy, Stony Brook University, Stony Brook, NY 11794-3800, USA}

$^2$\textit{Institute for Advanced Computational Science, Stony Brook University, Stony Brook, NY 11794-3800, USA}

$^3$\textit{Theoretical Division, Los Alamos National Laboratory, Los Alamos, NM 87545, USA}

$^4$\text{Center for Nonlinear Studies (CNLS), Los Alamos National Laboratory, Los Alamos, NM 87545, USA}

$^5$\textit{School of Applied and Engineering Physics, Cornell University, Ithaca, NY 14853, USA}

$^6$\textit{Center for Computational Quantum Physics, Flatiron Institute, 162 5$^{th}$ Avenue, New York, NY 10010, USA}

(Dated: \today)
\end{center}

\section{Water dissociation at $\text{SrTiO}_3$ [001] surfaces}
 %\vspace{-0.4 in}
%  \begin{figure}[h]
%  \includegraphics[width=4.in]{sto_waterfull.png}
%  \caption{From an equilibrated molecular dynamics simulation: (a) $(001)$ SrTiO$_3$ surfaces -- TiO$_2$ (left, in blue) and SrO (right, in green) -- interacting with a box of 64 water molecules. The number density distribution along the vertical $z-$direction for: (b) H$_2$O molecules and, (c) dissociated OH$^-$ species. The positions of TiO$_2$-- and SrO-- surfaces with respect to the box of water are shown in blue and green dashed lines respectively.}
%  \label{fig:sto_waterfull}
%  \end{figure}
 
 \begin{figure}[h]
 \includegraphics[width=7.in]{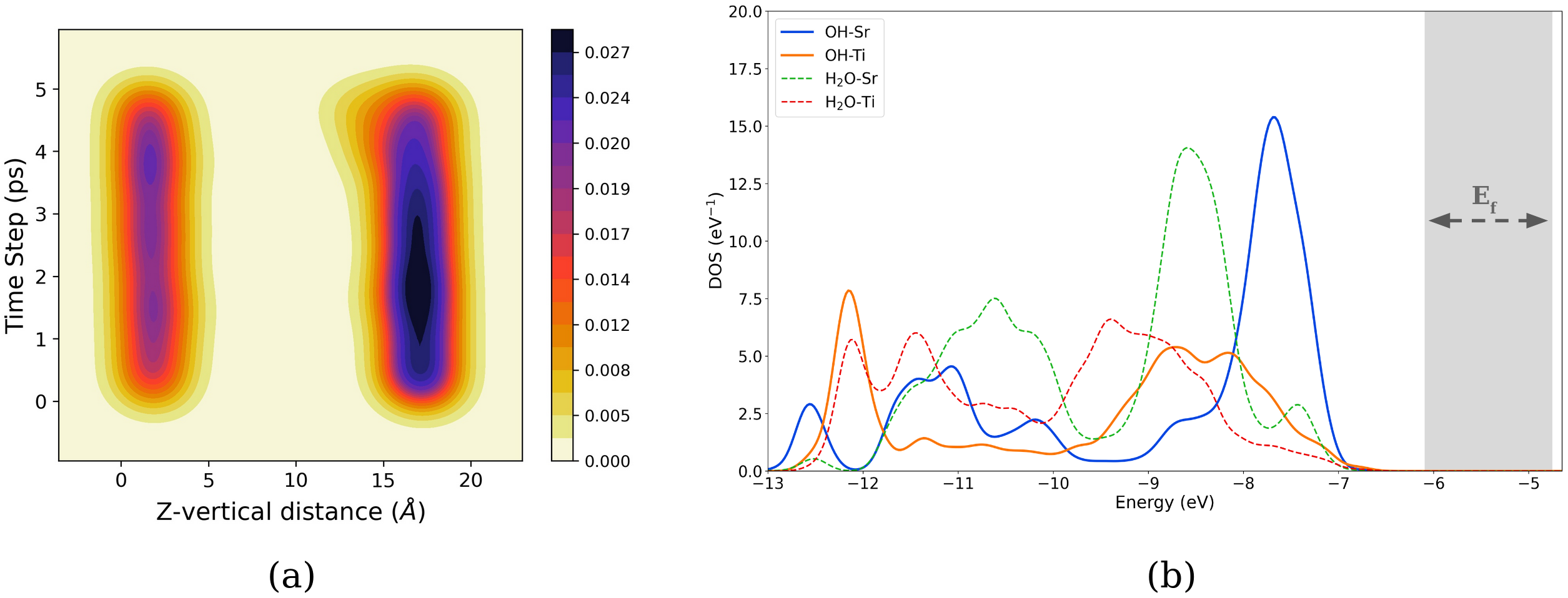}
 \caption{(a) Distribution of OH$^-$ along the vertical $z-$ direction for the equilibrated MD simulation time. The TiO$_2$-- and SrO-- surfaces are located on the left and right edges of the plot respectively. The colorbar indicates the density of OH at a given time and vertical position. (b) Density of states projected onto specific species and averaged over 20 different snapshots of the equilibrated MD trajectory. Solid lines indicate projection on surface hydroxyls (OH$^-$) after dissociation of the adsorbed water. Dashed lines indicate projection on the surface-adsorbed water (first monolayer).}
 \label{fig:oh_waterslab}
 \end{figure}

\section{[001] $\text{SrTiO}_3/\text{vacuum}$ symmetric-termination slabs}

The structures are geometrically relaxed within the DFT$+U$ ($U=4.45$ eV) formulation in \textsc{SIESTA} until the atomic forces are less than 0.01 eV/\AA.
In the following sections, electrostatic potential ($V_H$) is defined as the sum of the Hartree potential and the local pseudopotential at the grid used for DFT$+U$ simulations.
It differs from the total potential by the exchange-correlation potential term.
$V_H$ is averaged within the $xy-$plane for each atomic layer to give the `microscopic' potential ($\overline{V}_H$) along the $z-$direction.
$\overline{V}_H$ is then averaged out of the plane to yield the macroscopic potential $\left\langle V_H\right\rangle$.
The valence band maximum (HOMO) and conduction band minimum (LUMO) are derived from the eigenvalues given by DFT$+U$.

\newpage
 \begin{figure*}[h]
    \includegraphics[width=3.25in]{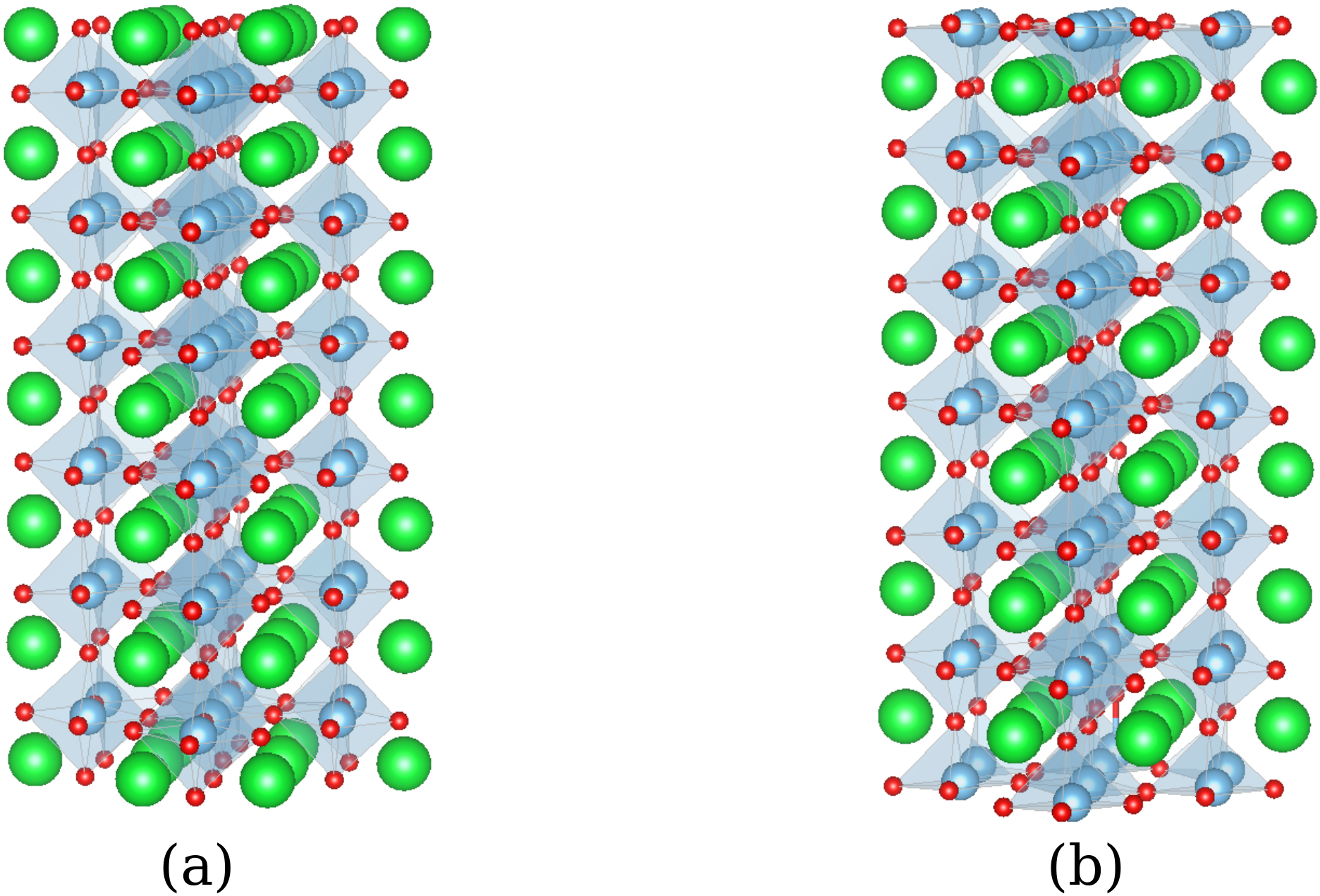}
    \caption{Structures of SrTiO$_3$ slabs: (a) $(2\sqrt{2}\times2\sqrt{2})$ SrO-terminated slab, (b) $(2\sqrt{2}\times2\sqrt{2})$ TiO$_2$-terminated slab. The lattice (box) parameters are chosen as: $a=b=11.059150$ \AA, $c=37.34$ \AA.}
    \label{fig:stoslabs}
 \end{figure*}

\subsection{[001] SrO-terminated slab}
\begin{figure*}[h]
    \includegraphics[width=6.in]{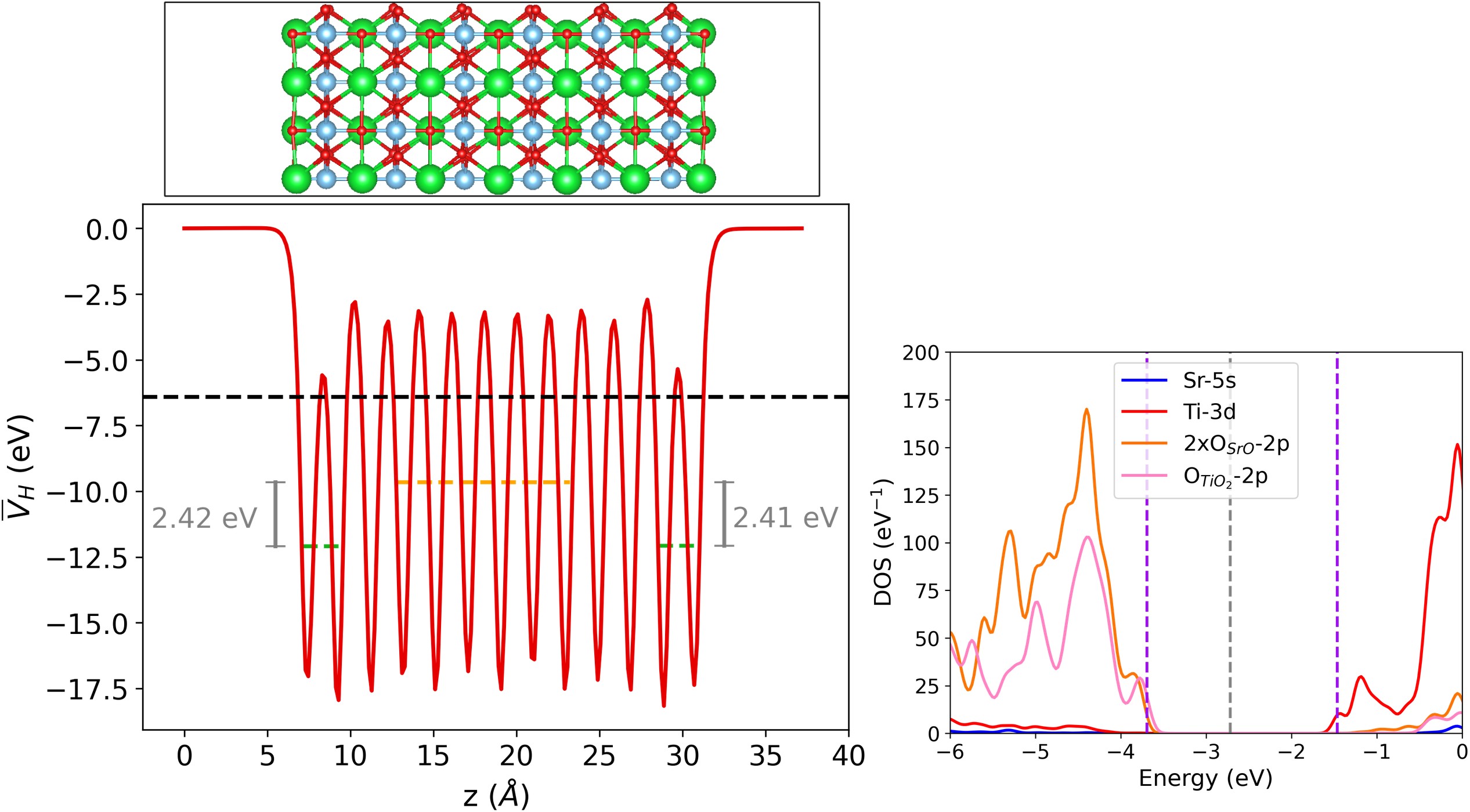}
    \caption{SrO-terminated $(001)$ SrTiO$_3$ slab: (left top) structure containing 7 SrO layers and 6 TiO$_2$ layers obtained with DFT$+U$, and (left bottom) planar-averaged electrostatic potential along $z-$ axis together with the macroscopic potentials computed in the bulk region of the slab $\left\langle V_{H}\right\rangle_{slab-bulk} = -9.659$ eV (orange dashed line) and at SrO- surfaces $\left\langle V_{H}\right\rangle_{SrO} = -12.088$ eV (green dashed lines). (Right) Corresponding projected density of states with Fermi level (grey dashed line), VBM or HOMO (left purple-dashed line) due to O$-2p$ states, CBM or LUMO (right purple-dashed line) due to Ti$-3d$ states.}
    \label{fig:srobands}
 \end{figure*}

The eigenvalues from DFT$+U$ are aligned to the vacuum level using the `zero' of the electrostatic potential.
The vacuum-aligned projected density of states are shown in Fig. \ref{fig:srobands}.
The valence and conduction band edges referenced to the vacuum are $E^{SrO}_\text{HOMO} = -3.693$ eV and $E^{SrO}_\text{LUMO} = -1.462$ eV respectively.

\subsection{[001] TiO$_2$-terminated slab}
\begin{figure*}[h]
    \includegraphics[width=6.in]{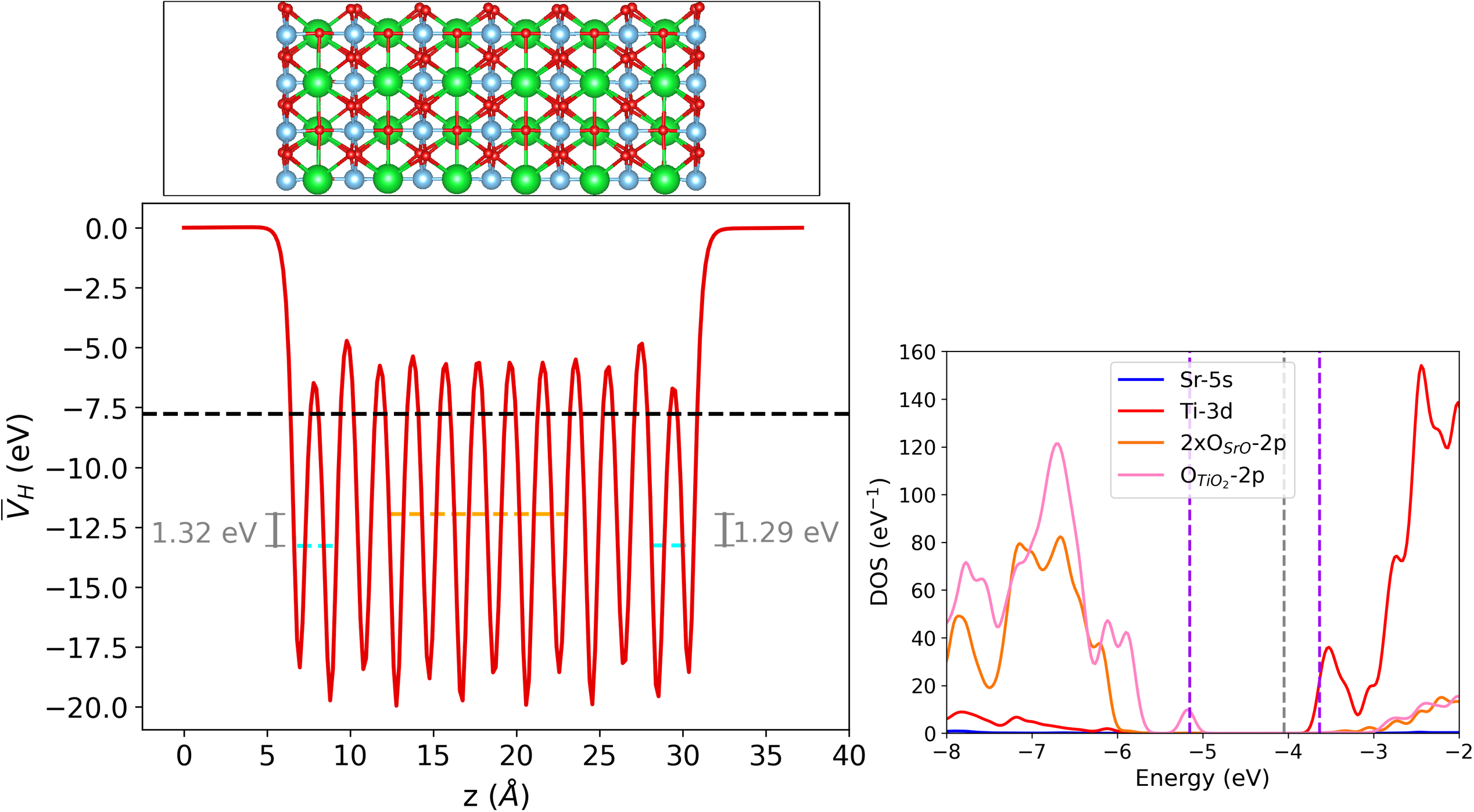}
    \caption{TiO$_2$-terminated $(001)$ SrTiO$_3$ slab: (left top) structure containing 7 TiO$_2$ layers and 6 SrO layers obtained with DFT$+U$, and (left bottom) planar-averaged electrostatic potential along $z-$ axis together with the macroscopic potentials computed in the bulk region of the slab $\left\langle V_{H}\right\rangle_{slab-bulk} = -11.956$ eV (orange dashed line) and at TiO$_2$- surfaces $\left\langle V_{H}\right\rangle_{TiO_2} = -13.276$ eV (cyan dashed lines). 
	(Right) Corresponding projected density of states with Fermi level (grey dashed line), VBM or HOMO (left purple-dashed line) due to O$-2p$ states, CBM or LUMO (right purple-dashed line) due to Ti$-3d$ states.}
    \label{fig:tio2bands}
 \end{figure*}

The zero of the projected density of states is shifted to match the vacuum level of the electrostatic potential, see Fig. \ref{fig:tio2bands}.
The band edges are hence located at $E^{TiO_2}_\text{HOMO} = -5.151$ eV and $E^{TiO_2}_\text{LUMO} = -3.631$ eV.

\section{Water $1b_1$ energy level alignment}

The water slab comprising 96 H$_2$O is approximately 20 \AA\, thick and alternates with 30 \AA\, region of vacuum. 
A Born-Oppenheimer molecular dynamics thermostatted to $T=330$ K is performed in order to simulate the dynamic properties of water.
From a 5 ps long equilibrated trajectory, 20 snapshots are selected and for each of these 20 geometries, the electrostatic potential and electronic density of states are computed using a GGA-type functional (vdW-BH).
The electrostatic potential and density of states are aligned to the vacuum level similar to the case of semiconductor band alignments discussed previously.

\begin{figure*}[h]
    \includegraphics[width=6.in]{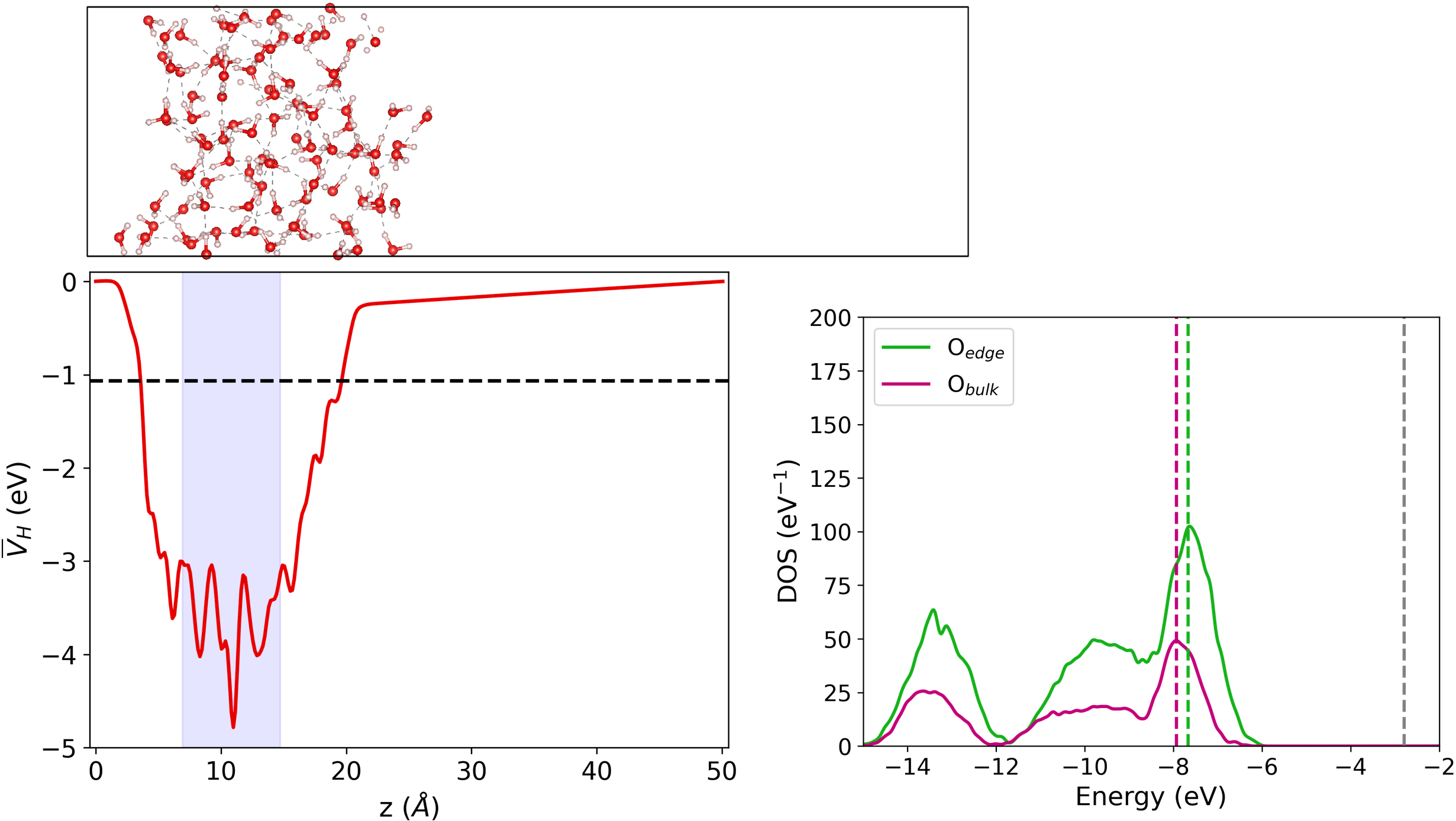}
    \caption{Electrostatic potential for the water-vacuum interface calculated using a water slab containing 96 H$_2$O and averaged over 20 snapshots of an MD simulation. The bulk potential is $\left\langle V_H \right\rangle_\text{water} = -3.62$ eV, indicated by the blue-shaded region. Vacuum-adjusted projected density of states for the bulk-like region of water: $E_{1b_1} = -7.93$ eV (pink dashed line).}
    \label{fig:waterbands}
 \end{figure*}

Figure \ref{fig:waterbands} shows the planar-averaged electrostatic potential ($\overline{V}_H$) along the $z-$direction and the density of states of Oxygen-$2p$ orbitals, averaged over 20 snapshots.
The macroscopic potential is determined for the bulk-like region of the water slab spanning an 8 \AA\, region (shown in blue) and the projected density of states are used to determine the position of the $1b_1$ peak in bulk water (pink dashed line).

\newpage
\section{Electrochemical energy level alignment at $\text{SrTiO}_3/$H$_2$O surfaces}

\begin{figure}[h]
 \includegraphics[width=6.in]{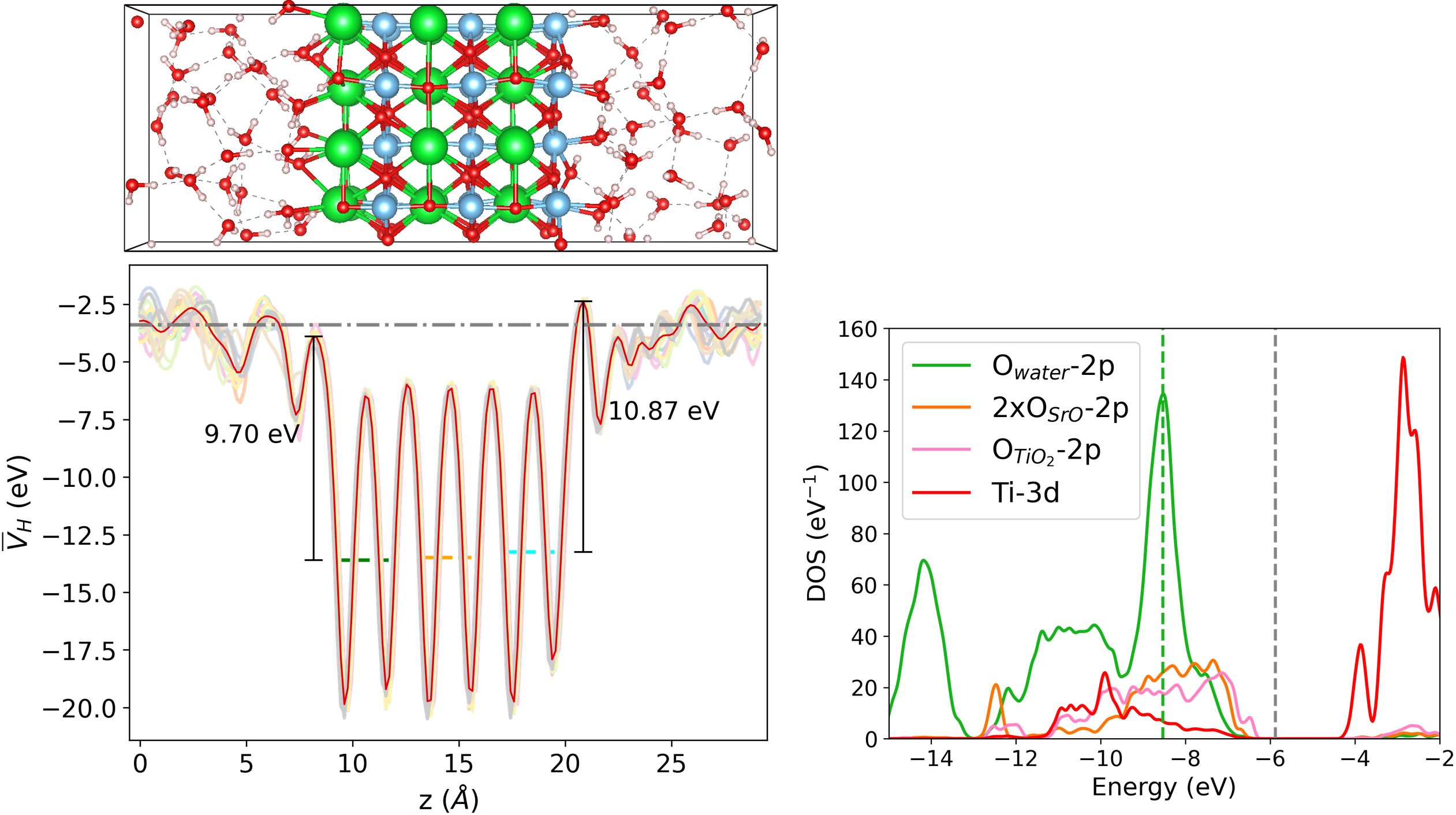}
 \caption{SrTiO$_3$/H$_2$O slab: (left top) structure from a snapshot of equilibrated MD trajectory, (left bottom) planar-averaged electrostatic potentials ($\overline{V}_H$) averaged over 20 snapshots (low-opacity colored lines) shown in red, (right) density of states projected onto oxygens of water, surface SrO and TiO$_2$ layers, and $3d$ orbitals of Ti.}
 \label{fig:water_stoalignment}
 \end{figure}

\begin{figure}[h]
    \includegraphics[width=5.25in]{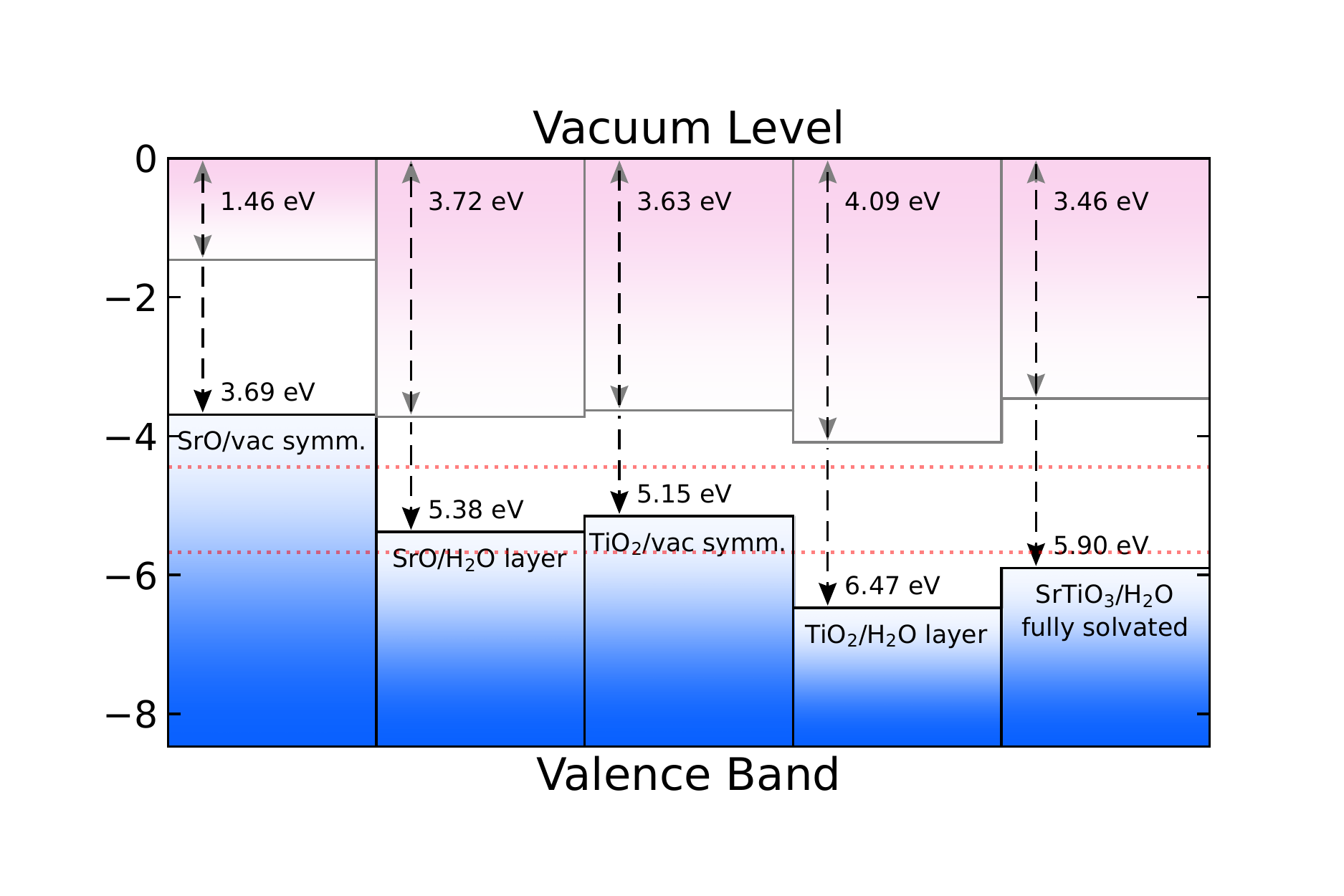}
    \caption{Band alignments for SrTiO$_3$/vacuum, SrTiO$_3$/H$_2$O monolayer and fully-solvated at DFT+U; U = 4.45 eV with a QZP basis set for oxygen. The red dotted lines indicate the water redox potentials referenced to the vacuum level (E$_{\text{H}^+/\text{H}_2} = -4.44$ eV, E$_{\text{O}_2/\text{H}_2\text{O}} = -5.67$ eV).}
    \label{fig:bandalignmentfull}
 \end{figure}

The electrostatic potentials are computed for 20 snapshots of SrTiO$_3$/H$_2$O system over a 5 ps time window of an equilibrated MD simulation.
These are first averaged over the $xy-$plane to obtain the microscopic potential along the $z-$direction for each individual snapshot, and then an average over all the snapshots is obtained to generate $\overline{V}_H$ shown in red in Fig. \ref{fig:water_stoalignment}.
$\overline{V}_H$ appears mostly flat in the bulk-like region of water ($0\leq z \leq 5$ \AA\, and $25\leq z\leq 30$ \AA, in Fig. \ref{fig:water_stoalignment}) implying that water screens the charge due to the asymmetric terminations of SrTiO$_3$.
From the density of states projected on the $2p$ orbitals belonging to Oxygen-water, we obtain the $1b_1$ peak at $E_{1b_1}=-8.53$ eV (green dashed line).
The highest occupied and lowest unoccupied energy levels are located at $-6.5$ eV and $-4.06$ eV respectively.
In order to align the energy levels across the semiconductor-water interfaces, we shift the $1b_1$ peak to the $1b_1$ level from the water-slab calculation shown in Fig. \ref{fig:waterbands}.
Thus, applying a rigid $\Delta=0.6$ eV shift to match the $1b_1$ levels of water, the valence and conduction band edges are found at $-5.9$ eV and $-3.46$ eV respectively.
The band alignment of the semiconductor band edges and the water redox potentials is shown in Fig. \ref{fig:bandalignmentfull}.
This was obtained at the GGA+U level using a quadruple-zeta basis set for Oxygens.

\newpage
\section{Proton-coupled electron transfer (PCET) at $\text{SrO}$- and $\text{TiO}_2$-terminated surfaces}
\begin{table*}[h]
\caption{\label{tab:stofull}Standard one-electron reduction potentials for PCET steps at SrO-terminated SrTiO$_3$-aqueous interface. The values are reported with respect to the normal hydrogen electrode (NHE) scale. The active sites $a$ and $b$ are shown in Fig. \ref{fig:sro_face}.}
\begin{tabular}{cccccccc} \hline\hline
Step & & Reaction on   & \multicolumn{2}{c}{Site $a$} &  & \multicolumn{2}{c}{Site $b$} \\
 & & SrO-termination &  $E^{\circ}$ (eV) & d$_{OO}$ (\AA) &  &  $E^{\circ}$ (eV) & d$_{OO}$ (\AA) \\ \hline
$1$ & & \multicolumn{1}{l}{(ii) $^{*}$O$^{\bullet -}$ \textcolor{red}{+ H$^{+}$ + e$^{-}$} $\to$ (i) $^{*}$OH$^{-}$} & $2.157$ & (ii) $2.655$ & & $2.102$ & (ii) $2.628$ \\ 
$2$ & &\multicolumn{1}{l}{(iii) $^{*}$OOH$^{-}$ \textcolor{red}{+ H$^{+}$ + e$^{-}$} $\to$ (ii) $^{*}$O$^{\bullet -}$ $+$ H$_2$O} & $-0.003$ & (iii) $1.476$ & & $-0.018$ & (iii) $1.490$ \\ 
$3$ & &\multicolumn{1}{l}{(iv) $^{*}$O$_2^{\bullet-}$ \textcolor{red}{+ H$^{+}$ + e$^{-}$} $\to$ (iii) $^{*}$OOH$^{-}$} & $0.478$ & (iv) $1.350$ & & $0.692$ & (iv) $1.356$ \\ 
$4$ & &\multicolumn{1}{l}{(v) $^{*}$OH$^{-}$ $+$ O$_2$ \textcolor{red}{+ H$^{+}$ + e$^{-}$} $\to$ (iv) $^{*}$O$_2^{\bullet-}$ $+$ H$_2$O} & $0.981$ & (v) $1.251$ & & $0.613$ & (v) $1.245$ \\ \hline
$1 + 2$ & & \multicolumn{1}{l}{$^{*}$OOH$^{-}$ \textcolor{red}{+ $2$H$^{+}$ + $2$e$^{-}$} $\to$ (i) $^{*}$OH$^{-}$ + H$_2$O} & $1.077$ & & & $1.042$ &  \\ 
$3 + 4$ & & \multicolumn{1}{l}{O$_2$ + $^{*}$OH$^{-}$ \textcolor{red}{+ $2$H$^{+}$ + $2$e$^{-}$} $\to$ $^{*}$OOH$^{-}$ + H$_2$O} & $0.730$ & & & $0.653$ & \\ \hline\hline
\end{tabular}	
\end{table*}

% \begin{table*}[h]
% \caption{\label{tab:tio2full}Standard one-electron reduction potentials for PCET steps at TiO$_2$-terminated SrTiO$_3$-aqueous interface. The values are reported with respect to the normal hydrogen electrode (NHE) scale. The active sites $c$, $d$ and $e$ are shown in Fig. \ref{fig:tio2_face}. Dagger ($\bm{\dagger}$) indicates the values for a loosely-bound O$-$OH$^{-}$ as the hydroperoxyl radical is \textit{not} formed in this case.}
% \begin{tabular}{cccccccccc} \hline\hline
% Step & & Reaction on   & \multicolumn{2}{c}{Site $c$} &  & \multicolumn{2}{c}{Site $d$} & \multicolumn{2}{c}{Site $e$} \\
%  & & TiO$_2$-termination &  $E^{\circ}$ (eV) & d$_{OO}$ (\AA) &  &  $E^{\circ}$ (eV) & d$_{OO}$ (\AA) & $E^{\circ}$ (eV) & d$_{OO}$ (\AA) \\ \hline
% $1$ & & \multicolumn{1}{l}{(ii) $^{*}$O$^{\bullet -}$ \textcolor{red}{+ H$^{+}$ + e$^{-}$} $\to$ (i) $^{*}$OH$^{-}$} & $1.854$ & (ii) $2.528$ & & $1.970$ & (ii) $2.549$ & $1.707$ & (ii) $2.613$ \\ 
% $2^{\bm{\dagger}}$ & &\multicolumn{1}{l}{(iii) $^{*}$OOH$^{-}$ \textcolor{red}{+ H$^{+}$ + e$^{-}$} $\to$ (ii) $^{*}$O$^{\bullet -}$ $+$ H$_2$O} & $1.929$ & (iii) $2.236$ & & $1.918$ & (iii) $2.408$ & $1.952$ & (iii) $2.208$\\ 
% & & \textcolor{red}{[NO $^{*}$OOH$^{-}$ formation]}\\
% \hline\hline
% \end{tabular}	
% \end{table*}
\begin{table*}[h]
\caption{\label{tab:tio2full}Standard one-electron reduction potentials for PCET steps at TiO$_2$-terminated SrTiO$_3$-aqueous interface. The values are reported with respect to the normal hydrogen electrode (NHE) scale. The active sites $c$, $d$ and $e$ are shown in Fig. \ref{fig:tio2_face}. Dagger ($\bm{\dagger}$) indicates the values for a loosely-bound O$-$OH$^{-}$ as the hydroperoxyl radical is \textit{not} formed in this case.}
\begin{tabular}{cccccccccc} \hline\hline
Step & & Reaction on   & \multicolumn{2}{c}{Site $c$} &  & \multicolumn{2}{c}{Site $d$} & \multicolumn{2}{c}{Site $e$} \\
 & & TiO$_2$-termination &  $E^{\circ}$ (eV) & d$_{OO}$ (\AA) &  &  $E^{\circ}$ (eV) & d$_{OO}$ (\AA) & $E^{\circ}$ (eV) & d$_{OO}$ (\AA) \\ \hline
$1$ & & \multicolumn{1}{l}{(ii) $^{*}$O$^{\bullet -}$ \textcolor{red}{+ H$^{+}$ + e$^{-}$} $\to$ (i) $^{*}$OH$^{-}$} & $1.884$ & (ii) $2.528$ & & $1.970$ & (ii) $2.549$ & $1.707$ & (ii) $2.613$ \\ 
$2^{\bm{\dagger}}$ & &\multicolumn{1}{l}{(iii) $^{*}$OOH$^{-}$ \textcolor{red}{+ H$^{+}$ + e$^{-}$} $\to$ (ii) $^{*}$O$^{\bullet -}$ $+$ H$_2$O} & $1.917$ & (iii) $2.236$ & & $1.918$ & (iii) $2.408$ & $1.952$ & (iii) $2.208$\\ 
& & \textcolor{red}{[NO $^{*}$OOH$^{-}$ formation]}\\
\hline\hline
\end{tabular}	
\end{table*}

\begin{figure}[h]
\begin{subfigure}{0.45\textwidth}
    \includegraphics[width=2.in]{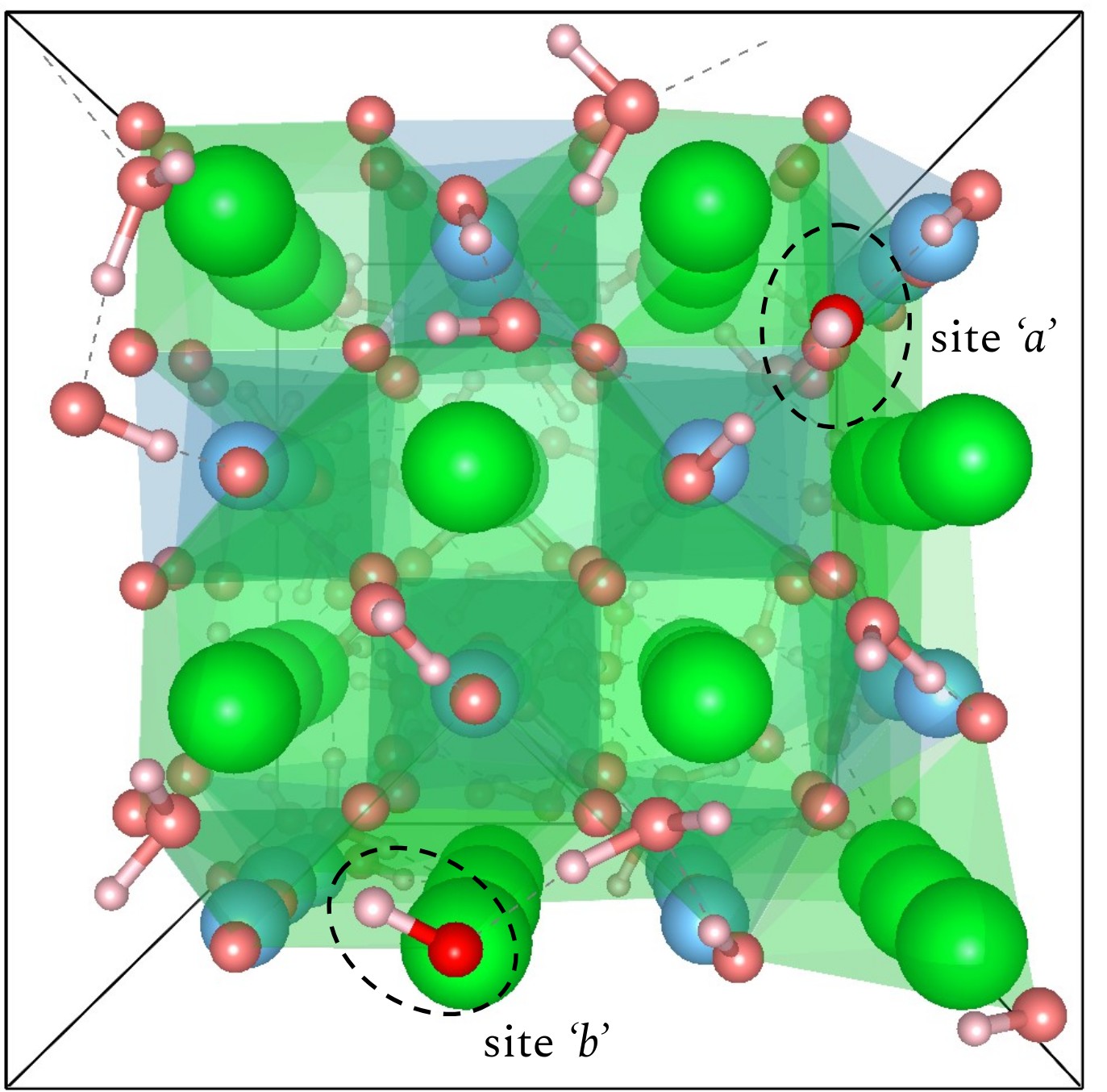}
    \caption{}
    \label{fig:sro_face}
\end{subfigure}
\begin{subfigure}{0.45\textwidth}
    \includegraphics[width=2.in]{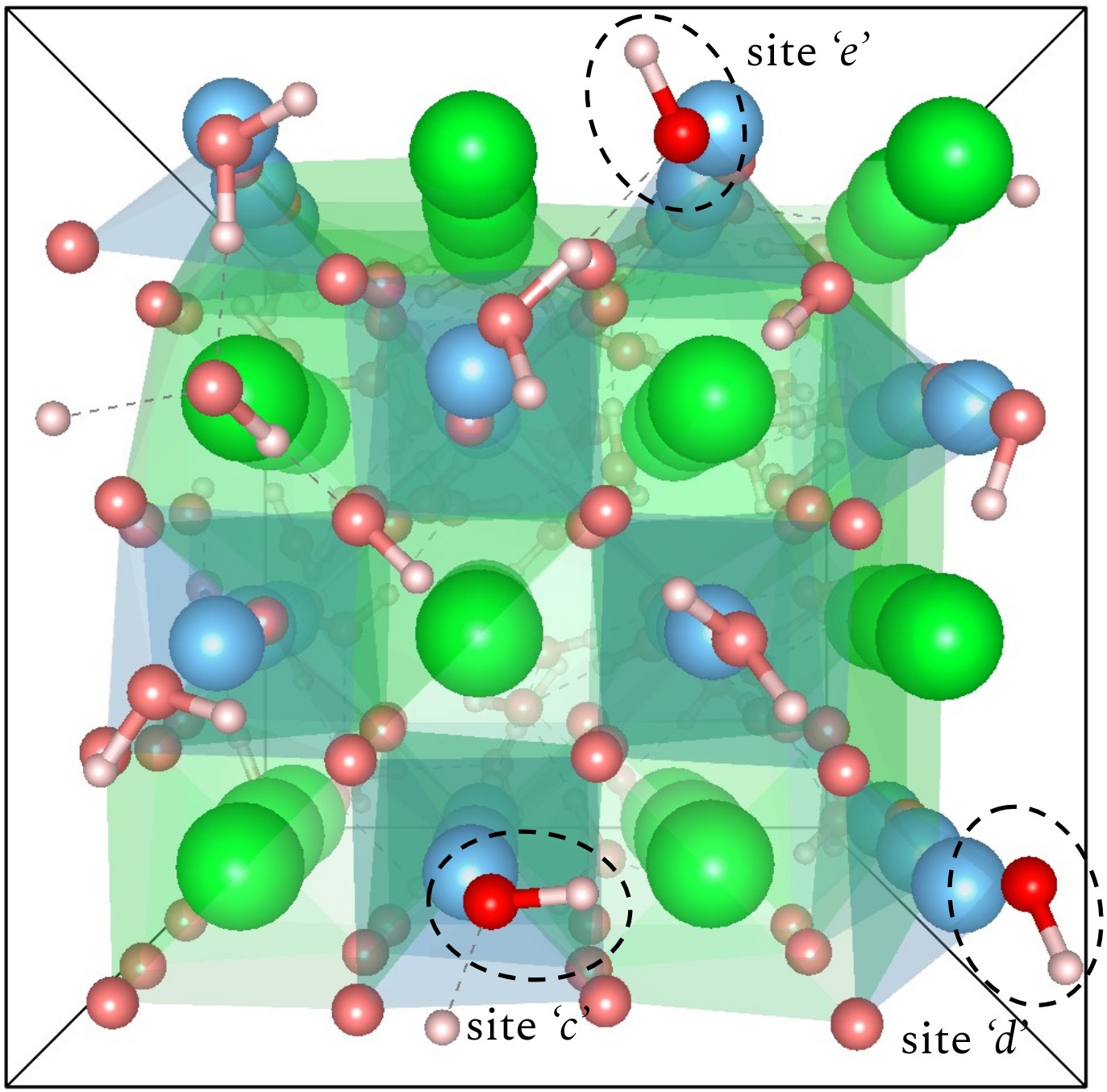}
    \caption{}
    \label{fig:tio2_face}
\end{subfigure}
\caption{SrTiO$_3$/Water interface showing dissociated and physisorbed water. The surface-hydroxyl species relevant for PCET calculations are highlighted for (a) SrO-surface and (b) TiO$_2$-surface termination.}
\end{figure}

% needed for \newpage to work correctly
% in the absence of a more elegant solution
% \textcolor{white}{Dummy text}\\
% \textcolor{white}{Dummy text}
%
% \newpage

%%%%%%%%%%%%%%%%%%%%%%%%%%%%%%%%%%%%%%%%%%%%%%%%%%%%%%%%%%%%%%%%%%%
%%%              SUPPLEMENTAL  MATERIAL  ENDS HERE          %%%
%%%%%%%%%%%%%%%%%%%%%%%%%%%%%%%%%%%%%%%%%%%%%%%%%%%%%%%%%%%%%%%%%%%

\end{document}